\documentclass[journal]{IEEEtran}
\usepackage{xcolor,soul,framed} 
\usepackage{subfigure}

\usepackage[numbers,sort&compress]{natbib}
\colorlet{shadecolor}{yellow}
\usepackage[pdftex]{graphicx}
\usepackage{booktabs}
\graphicspath{{../pdf/}{../jpeg/}}
\DeclareGraphicsExtensions{.pdf,.jpeg,.png}
\usepackage[cmex10]{amsmath}
\usepackage{array}
\usepackage{amsmath}
\usepackage{amsfonts}
\usepackage{multirow} 
\usepackage{mdwmath}
\usepackage{mathabx}
\usepackage{mdwtab}
\usepackage{eqparbox}
\usepackage{url}
\usepackage{color}
\usepackage{algorithm}
\usepackage[justification=centering]{caption}
\usepackage{caption}
\usepackage{graphicx}
\usepackage{float} 
\usepackage{caption}
\usepackage{graphicx, subfig}
\usepackage{subfigure}
\usepackage{harpoon}
\usepackage{algorithm}
\usepackage{amsmath}        
\usepackage{amssymb}        
\usepackage{algorithm}      
\usepackage{algpseudocode}  

\begin{document}


\bstctlcite{IEEEexample:BSTcontrol}
    \title{Goal-oriented Semantic Communications for Metaverse Construction via Generative AI and Optimal Transport}
    \author{Zhe Wang, Nan Li,
 Yansha Deng, ~\IEEEmembership{Senior Member,~IEEE, }
 and A. Hamid Aghvami, ~\IEEEmembership{Life Fellow,~IEEE }
 \thanks{Z. Wang, N. Li, Y. Deng, and A. Hamid Aghvami (Emeritus Professor) are with the Department of Engineering, King’s College
London, Strand, London WC2R 2LS, U.K. (e-mail: tylor.wang@kcl.ac.uk; nan.3.li@kcl.ac.uk
yansha.deng@kcl.ac.uk; hamid.aghvami@kcl.ac.uk)
 }
 


}



\maketitle

\begin{abstract}
The emergence of the metaverse has boosted productivity and creativity, driving real-time updates and personalized content, which will substantially increase data traffic. However, current bit-oriented communication networks struggle to manage this high volume of dynamic information, restricting metaverse applications interactivity.
To address this research gap, we propose a goal-oriented semantic communication (GSC) framework for metaverse. Building on an existing metaverse wireless construction task, our proposed GSC framework includes an hourglass network-based (HgNet) encoder to extract semantic information of objects in the metaverse; and a semantic decoder that uses this extracted information to reconstruct the metaverse content after wireless transmission, enabling efficient communication and real-time object behaviour updates to the scenery for metaverse construction task.
To overcome the wireless channel noise at the receiver, we design an optimal transport (OT)-enabled semantic denoiser, which enhances the accuracy of metaverse scenery through wireless communication. 
Experimental results show that compared to the conventional metaverse construction, our proposed GSC framework significantly reduces wireless metaverse construction latency by 92.6\%, while improving metaverse object status accuracy and viewing experience by 45.6\% and 44.7\%, respectively.
\end{abstract}


%
\IEEEpeerreviewmaketitle

\section{INTRODUCTION}
The concept of the metaverse has emerged as a comprehensive extension of the digital universe, encompassing various applications such as real-world scenario construction and complex simulations \cite{wang2022survey}. 
Most previous researches in metaverse focused on achieving precise rendering of visual content from physical counterparts through wired and wireless communication \cite{zhao2022metaverse}. 
The metaverse construction task serves as a foundational step for other applications, such as BMW's intelligent virtual factories, where production efficiency can be analyzed and optimized through metaverse simulation \cite{hu2023research}.
The goal of these metaverse construction task is to accurately transmit different types of metaverse data using bit-oriented communication methods. 
However, supporting real-time metaverse interactions and high-quality rendering typically requires a bandwidth of approximately 5.6 Gbps for raw graphic data downloads \cite{dong2022metaverse}. This requirement far exceeds the global average 5G wireless download speed of 160 Mbps \cite{ng2022metaverse}, posing a bottleneck for metaverse applications.

To generate a metaverse scenery, high-dimensional data like point clouds \cite{huang2023iscom} and meshes \cite{singh20238} can be utilized to capture more comprehensive information, including spatial positions, color attributes, and depth information, that can provide a richer interaction for clients.
Transmitting raw point clouds or meshes in metaverse construction requires a large amount of bandwidth. For example, a 30 FPS point cloud video generates approximately 2.06 Gb of raw data per second \cite{huang2023iscom}.
Given these high data rates required to transmit these point clouds or meshes, current metaverse construction tasks often rely on high-resolution images \cite{wong2022360} and video inputs for processing and rendering into virtual environments, rather than using raw metaverse data formats.
However, rendering the metaverse from images and videos is both time and resource intensive, particularly for large-scale scenes that are challenging to render in real time. Previous research has shown that creating an interactive metaverse scene demands high computational resources, generally requiring over two days of training on two Tesla V100 GPUs \cite{jing2023reconstruction}. Small amount of inaccuracies and delays in data transmission can undermine the metaverse reliability and realism, thereby degrading the user viewing experience.

To address the communication challenges posed by data-intensive metaverse applications, the concept of semantic communication has emerged as a potential solution. 
Unlike traditional communication methods, semantic communication updates the static knowledge, which includes shared information and maintains consistency between the transmitter and receiver \cite{chaccour2024less}, allowing for meaningful information transmission and thus reducing the demand for high bandwidth \cite{xu2023knowledge}.
Recent research has explored the application of semantic communication, across image-centric contexts \cite{cheng2024magicstream} and video-specific contexts \cite{li2024video} through the design of semantic encoders and decoders.
For image communication task, \cite{yamamoto2024deep} presented a JSCC-enabled semantic encoder-decoder framework that enhances image construction accuracy by mapping key features as semantic information and using image overlap patches.
However, JSCC-enabled image semantic communication operates as end-to-end deep learning frameworks, jointly optimizing source and channel coding through deep neural networks, making it challenging to apply in scenarios where training data may not fully capture all features of the transmission signals and wireless channel \cite{jain2005introduction}.

For video transmission tasks, researchers primarily rely on frame interpolation techniques to generate videos by processing image sequences frame-by-frame \cite{samarathunga2024semantic, bao2023mdvsc, wang2022wireless}. While frame interpolation effectively smooths gaps between frames, this approach has inherent limitations in generating cohesive video sequences due to its lack of unified perspective on frame continuity. As a result, these methods often struggle to produce smooth, realistic video, particularly when complex motion or scene consistency is required.  Recently, video generation-enabled semantic decoders, such as the text-to-video model \cite{cho2024sora}, have emerged as a promising solution to address these limitations through semantic interpretation. These models are designed to create more contextually coherent frames by understanding the high-level content and structure of scenes. However, fully realizing a robust text-to-image-to-video pipeline remains challenging, particularly in maintaining consistent object appearance and placement within static scenes across entire video sequences \cite{liu2024sora}. This challenge is especially critical in real-time applications like the metaverse, where generating coherent 2D video sequences within stable 3D virtual environments is crucial. Metaverse construction shows significant potential for enhancing video generation by providing a spatially consistent 3D framework, which could serve as a foundation for generating more coherent video content. Advancing these capabilities could open new possibilities for creating immersive and stable video experiences within 3D virtual environments.

To construct metaverse scenery, generative AI frameworks such as Stable Diffusion (SD) \cite{guo2023animatediff} and Neural Radiance Field (NeRF) \cite{mildenhall2021nerf} have demonstrated remarkable capabilities in synthesizing customized images and 3D models through text-driven prompts and control mechanisms, respectively. These frameworks have significantly advanced the state-of-the-art in computational content generation for virtual environments.
Specifically, \cite{zhang2024tale} leveraged SD-based postprocessing to enhance semantic object matching and spatial relationship modeling from different viewing angles, demonstrating improved accuracy in object generation and positioning within synthesized scenes. 
Similarly, \cite{xu2022point} employed NeRF for 3D scene reconstruction, utilizing objects structural relationships for optimized sampling in large-scale scene rendering, thereby enabling efficient representation of complex virtual environments.
However, these generative AI frameworks primarily emphasize custom and reliable image generation, with limited consideration on the impact of the practical constraints of wireless transmission systems on content generation. 
Especially the transmission errors and delays caused by fading and noise in wireless channels highlight a critical research gap in metaverse content delivery.
This impact of wireless channel becomes particularly critical for metaverse applications with the real-time requirements and reliability demands in real wireless communication environments.

To address the impact of noise and channel fading on semantic information transmission, machine learning approaches, have demonstrated promising denoising capabilities at the receiver. However, due to inherent unexplainable issues in machine learning, mathematical optimization methods, specifically Optimal Transport (OT) theory, have exhibited superior performance in optimizing large-volume data distribution \cite{sejourne2023unbalanced}.
OT, originally proposed to minimize mass transfer costs between probability distributions, provides robust objective functions and ensures consistency in data while addressing resource constraints. This makes it particularly well-suited for applications involving high-dimensional data \cite{martyushev2023stochastic}.
Recently, OT has gained significant research attention in image processing applications, where \cite{sana2023semantic} implemented OT in multi-user semantic communication systems through distinct semantic decoders and channel equalizers to resolve language ambiguities and semantic mismatches, and \cite{Liu2020Semantic} formulated semantic correspondence as an OT problem to align style disparities across semantically similar images, establishing dense correspondences at the semantic level. 
However, existing research on OT mainly focused on preprocessing or postprocessing at the transmitter or receiver, lacking comprehensive consideration of wireless channel fading impacts on semantic communication. 
This research gap becomes particularly evident in high-dimensional datasets, such as those encountered in metaverse applications, where complex distributional characteristics present unique challenges and opportunities for further exploration.

To address the above limitations, we propose a goal-oriented semantic communication framework (GSC). 
Unlike a wireless metaverse construction framework with image transmission, our proposed GSC framework extracts both 1D and 2D semantic information from high dimensional metaverse data, incorporating a semantic denoising module to achieve lower bandwidth usage while maintaining communication accuracy. 
The contributions of this paper are summarized as follows:
\begin{itemize}
    \item We propose a goal-oriented semantic communication (GSC) framework that allows users to customize content based on their innovations and inspirations, enabling changes in color, style, and object status updates within the metaverse construction task. The proposed GSC framework includes a semantic encoder to extract key points from metaverse scenery, an OT-enabled semantic denoiser to optimize semantic noise, and a semantic decoder with stable diffusion and NeRF for metaverse construction.
    \item We propose a novel OT-enabled semantic denoiser that consists of a semantic selective correction algorithm to effectively reduce noise in the received semantic information. This is achieved by minimizing the difference between the distributions of the denoised data points and the originally transmitted data points. 
    \item We conduct extensive experiments to demonstrate the significant improvements of our proposed GSC framework over the conventional metaverse construction. Specifically, our framework achieved improvements over conventional metaverse construction methods, including a 45.6\% increase in metaverse status accuracy, a 44.7\% enhancement in viewing experience, and a 92.6\% reduction in transmission latency.
\end{itemize}
The rest of the paper is organized as follows.
Section II presents the system model and problem formulation.
Section III describes the different modules of the proposed GSC framework.
Section IV outlines the design principles of the OT-enabled semantic denoiser.
Section V discusses the evaluation metrics and experimental results.
Finally, Section VI concludes the paper.

\begin{figure}[t]
  \centering
  \includegraphics[width=.45\textwidth]{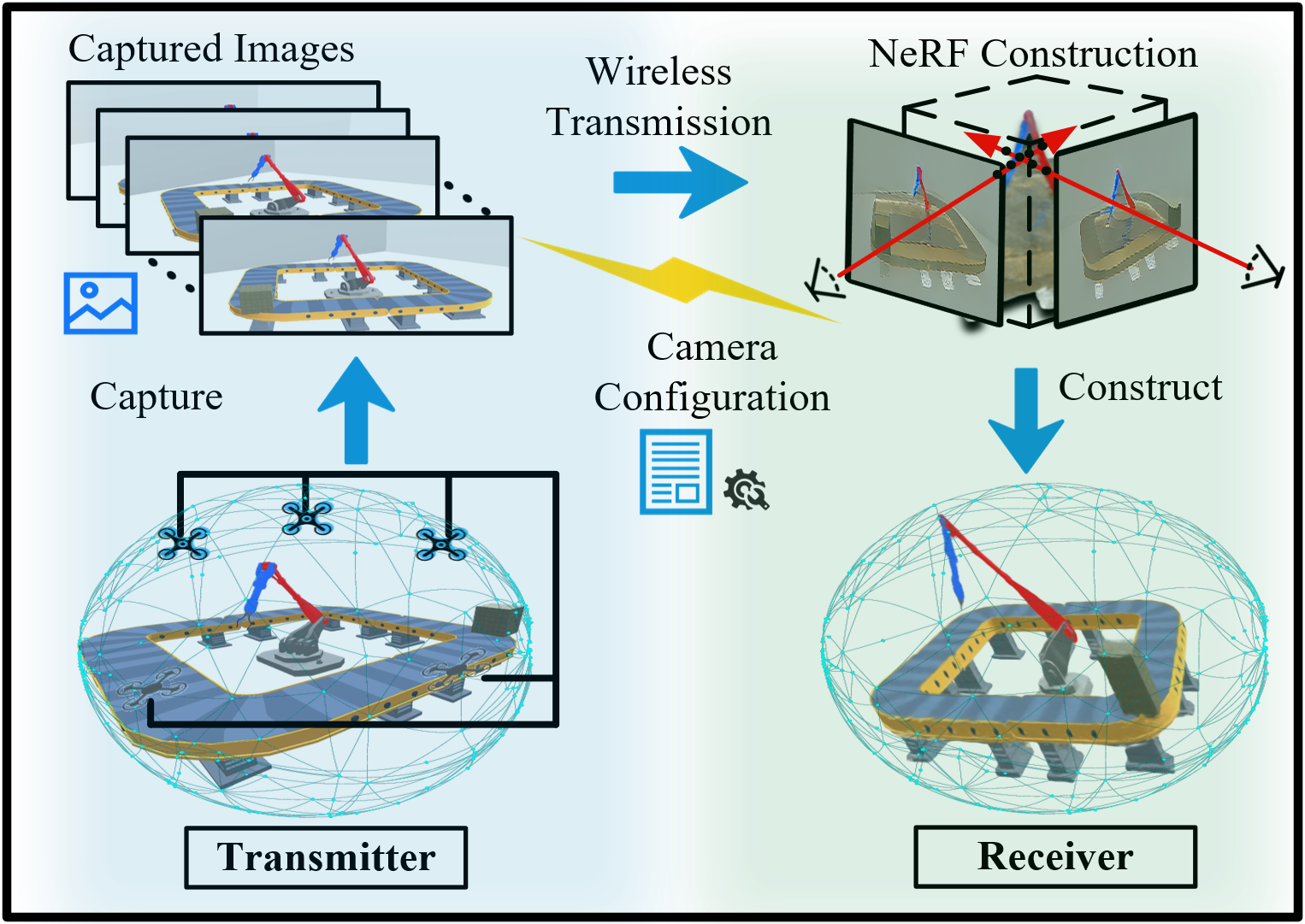} 
  \caption{Metaverse Construction Task} 
  \label{traditionalimagecom} 
\end{figure}

\section{SYSTEM MODEL AND PROBLEM FORMATION}
In this section, we provide an overview of the traditional wireless metaverse construction tasks and the problem formulation. As shown in Fig. \ref{traditionalimagecom}, the process of traditional metaverse construction includes image capturing, image-based wireless transmission, and metaverse construction based on image input.


\subsection{System Model}
We consider a metaverse construction task in an industrial factory scenario \cite{xu2022point}, aiming to replicate a physical factory and its operational status within the metaverse.
The factory scenery features a stationary conveyor belt as a stable object, with several moving elements like a box traveling along the belt and a robotic arm operating in the middle, as plotted in Fig. \ref{traditionalimagecom}.
In this scenery, a set of UAVs is evenly positioned at fixed area defined by the dimensions of length ($\text{L}$), width ($\text{W}$), and height ($\text{H}$), forming a structured metaverse environment.
At each time slot $t$, each UAV captures an image from its specific orientation $\theta$. The configuration set $\mathcal{C}$ for the UAV cameras encompasses several fixed parameters, which are defined as
\begin{equation}
\label{camera_information}
\mathcal{C}=\{r_{\theta}, (f_x, f_y), (c_x, c_y)\},
\end{equation}
where the homogeneous transformation rotation matrix $r_{\theta}$ represents the orientation and position of the camera coordinate system relative to the origin of the world coordinate system. 
The parameters ($f_x$, $f_y$) represent the focal lengths, indicating the camera’s magnification level along each axis.
We denote $(c_x, c_y)$ as the principal point offsets, which specify the image center relative to the sensor’s coordinate system. The set of captured images at time slot $t$ is denoted as $\mathbf{V}_t$, and can be represented as
\begin{equation}
\mathbf{V}_t=\left[\mathbf{I}_1, \cdots, \mathbf{I}_{\mathrm{N}_{\mathrm{u}}}\right]^{\mathrm{T}}, \ \mathbf{I}_i \in \mathbb{R}^{\text{H} \times \text{W}},
\end{equation}
where $\text{N}_{\text{u}}$ denotes the total number of UAVs, and $\mathbf{I}_i$ represents the RGB image matrix captured by the $i$-th UAV. Each captured image is in RGB format and shares the same resolution across all UAVs.

In the wireless transmission process, a Rayleigh fading channel is introduced to model signal strength fluctuations caused by environmental factors such as mobility, multipath propagation, and unpredictable conditions. These fluctuations are represented by the channel matrix $\mathbf{H}$, which is composed of individual fading parameters $h_i$ defined as
\begin{equation}
h_i=\sqrt{n_{1 i}^2+n_{2 i}^2}, \quad n_{1 i}, n_{2 i} \sim \mathcal{N}\left(0, \frac{1}{2}\right),
\end{equation}
where $n_{1i}$ and $n_{2i}$ are independent and identically distributed Gaussian random variables. 
During transmission, the data matrix $\mathbf{V}_t$ is applied through the channel matrix $\mathbf{H}$ and affected by additive noise, resulting in the received data matrix $\mathbf{V}_t^{\prime}$, which can be expressed as
\begin{equation}
\mathbf{V}_t^{\prime}=\mathbf{H} \otimes \mathbf{V}_t + \mathbf{N},
\end{equation}
where $\otimes$ denotes convolution, and $\mathbf{N}$ represents the additive noise, which has the same size as $\mathbf{V}_t$.


At the receiver, the received image set $\mathbf{V}^{\prime}_t$, along with the UAV camera parameters $\mathcal{C}^{\prime}$, is used to construct the metaverse scenery. This process results in a metaverse data representation, such as point cloud, denoted as $\mathbf{P}_{\text{r}}$, formulated as
\begin{equation} 
\label{construct} \mathbf{P}_{\text{r}}=[\overrightharp{v}_{1}, \cdots, \overrightharp{v}_{\text{N}_{\text{r}}} ]^\text{T}=\mathcal{R}\left(\mathbf{V}^{\prime}_t, \mathcal{C}^{\prime}, \delta_{\text{r}} \right), 
\end{equation} 
where the variable $\delta_{\text{r}}$ represents NeRF algorithm parameters, $\text{N}_{\text{r}}$ represents the number of points in the metaverse scenery, and each point vector $\overrightharp{v}_{i}$ can be represented as
\begin{equation}
\overrightharp{v}_{i}=(\overrightharp{l}_i, \overrightharp{c}_i)=(l_\text{x}, l_\text{y}, l_\text{z}, c_\text{r}, c_\text{g}, c_\text{b}),
\end{equation}
where the $\overrightharp{l}_i$ and $\overrightharp{c}_i$ represent the three-dimensional location and RGB color of point, respectively.

\begin{figure*}[t]
  \centering
  \includegraphics[width=.9\textwidth]{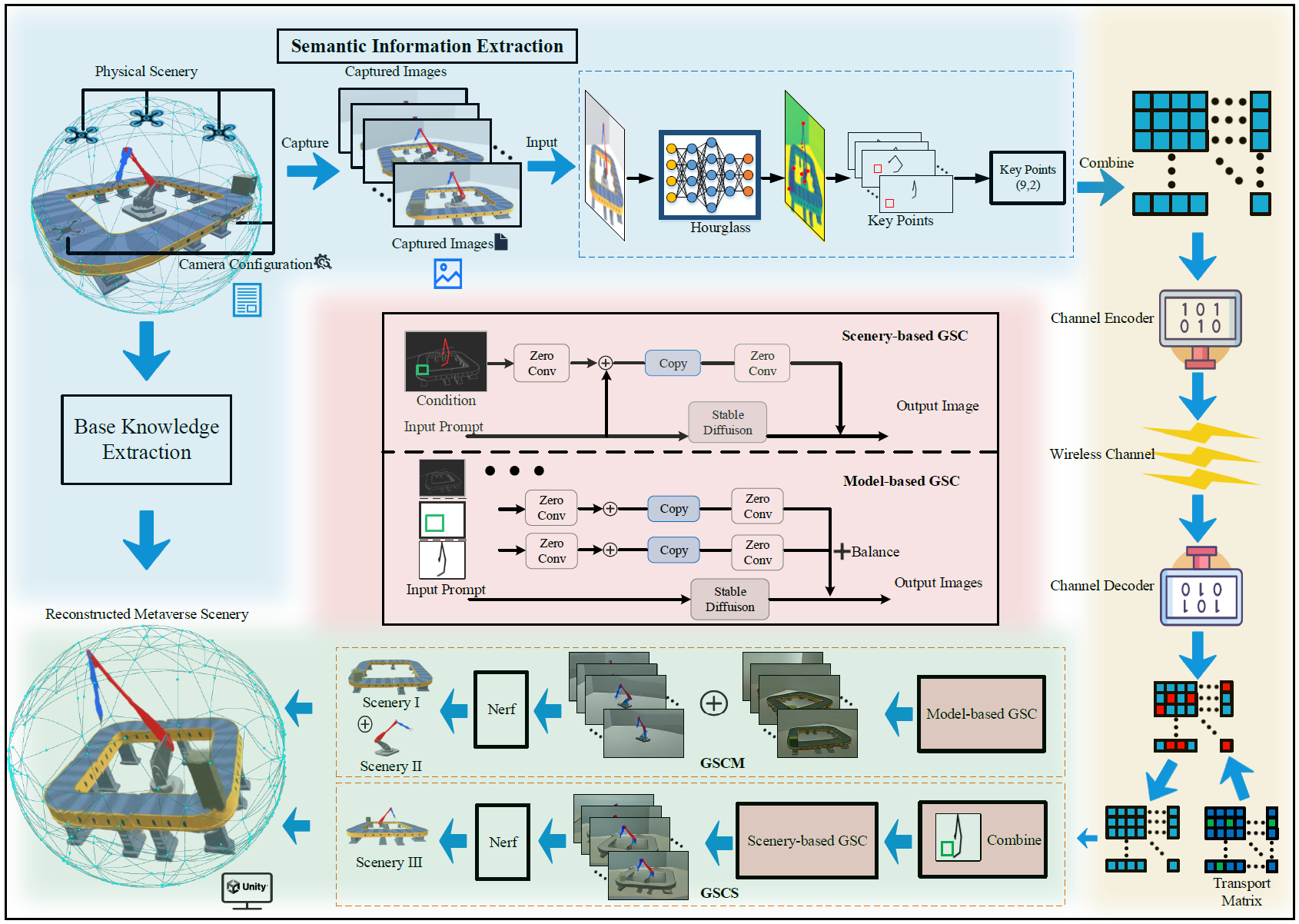} 
  \caption{Goal-oriented Semantic Communication Framework for the Metaverse Construction} 
  \label{metaverse_construction} 
\end{figure*}

\subsection{Problem Formation}
The goal of metaverse construction is to maintain consistent scenery between the transmitter and receiver. Previous research has shown that objects in point cloud digital twins are easier for performance evaluation compared to other formats. 
Notably, color and style enhance object recognition in point clouds primarily when motion is involved, whereas point clouds themselves are efficient for 3D object detection \cite{bremner2022impact}. Evaluating metaverse construction based on point clouds can effectively reflect both detection accuracy and the viewing experience in metaverse applications.
This demonstrates that to ensure an optimal viewing experience on the client side based on the transmitted data, it is essential for the point clouds at the transmitter and the receiver to closely match each other.
To achieve this consistency, we aim to minimize the geometry difference between the point cloud representation $\mathbf{P}_{\text{t}}$ at the transmitter and $\mathbf{P}_{\text{r}}$ at the receiver, using a modified chamfer distance measure $\mathcal{C}(\cdot)$, which is calculated as
\begin{equation}
\begin{aligned}
\label{objective_fun}
\mathcal{P}: &\min _{\mathbf{P}_{\mathrm{r}}}
\mathcal{C} \Big( \mathbf{P}_{\text{t}}, \ \mathbf{P}_{\text{t}}\Big)\\
= & \min _{\left\{r_{\theta},  \delta_{\text{r}}\right\}} \frac{1}{\left|\mathbf{P}_{\mathrm{t}}\right|} \sum_{\overrightharp{v} \in \mathbf{P}_{\mathrm{t}}} \min _{\overrightharp{v}^{\prime} \in \mathbf{P}_{\mathrm{t}}}\left\|\overrightharp{v}-\overrightharp{v}^{\prime}\right\|^2 +\\
& \frac{1}{\left|\mathbf{P}_{\mathrm{r}}\right|} \sum_{\overrightharp{v}^{\prime} \in \mathbf{P}_{\mathrm{r}}} \min _{\overrightharp{v} \in \mathbf{P}_{\mathrm{r}}}\left\|\overrightharp{v}^{\prime}-\overrightharp{v}\right\|^2,
\end{aligned}
\end{equation}


where $|\mathbf{P}|$ represents the number of points in the point cloud $\mathbf{P}$, the objective is to minimize the distance between the transmitter and receiver by aligning their respective metaverse data representations as closely as possible.

\subsection{Evaluation Metrics}
The overall goal of metaverse construction task, as shown in Eq. (\ref{objective_fun}) is to recover the scenery for better metaverse status and clients viewing experience. To achieve this, we use different metrics to evaluate the entire virtual scenery, which includes key point error (KPE), point-to-point (P2Point), and transmission latency.

\textbf{P2Point} \cite{mekuria2016evaluation}: To evaluate the viewing experience of clients in metaverse, the P2Point metric is employed to assess the generated scenery from a $360^{\circ}$ viewing angles, comparing the geometry difference between the point cloud data at transmitter $\mathbf{P}_{\text{t}}$ and the point cloud data generated at receiver $\mathbf{P}_{\text{r}}$.
The P2Point error calculation can be expressed as
\begin{equation}
\text{P2Point}= \max \left({d}_{\text{rms}}{\left(\mathbf{P}_{\text{t}}, \mathbf{P}_{\text{r}}\right)}, {d}_\text{rms}{\left(\mathbf{P}_{\text{r}}, \mathbf{P}_{\text{t}}\right)}\right),
\end{equation}
where the function ${d}_{\text{rms}}$ is the root mean square error between two point cloud.

\textbf{KPE}:
The KPE is used to estimate and evaluate the metaverse object status key points error between the transmitted images and received images, which can be expressed as
\begin{equation}
\label{kpe_calculate}
\begin{aligned}
& \text { KPE }=\frac{1}{N} \sum_{i=1}^{N} \sqrt{{|\overrightharp{K}_i-\overrightharp{K}^{'}_{i}|}^2},
\end{aligned}
\end{equation}
where the $\overrightharp{K}_i$ and $\overrightharp{K}^{'}_i$ represent the three dimensional position value of key points at the transmitter and the receiver respectively, $N$ represents the total number of points in each image.

\textbf{Latency}: Latency is a critical metric in metaverse applications. The transmission latency of the metaverse construction task can be divided into different components, including semantic information extraction time  $T_\text{s}$, wireless communication time $T_\text{w}$, OT selective correction time $T_\text{o}$, and image generation time $T_\text{g}$. The combination of all these times results in the transmission delay of the metaverse application, which can be expressed as
\begin{equation}
\label{time_delay}
\text{L}=T_\text{s}+T_\text{w}+T_\text{o}+T_\text{g},
\end{equation}
by analyzing and optimizing each component of the transmission latency, we can justify and indicate the efficiency of our proposed framework.

\section{Goal-oriented Semantic Communication}
In this section, we present an overview of the goal-oriented semantic communication (GSC) framework for wireless metaverse reconstruction, as shown in Fig. \ref{metaverse_construction}. The framework consists of four main modules: the knowledge base extraction module, which gathers and updates essential static knowledge base; the semantic encoder, which takes images as input and generates 1D semantic information as output; the wireless communication module; and the semantic decoder, which  reconstructs the metaverse scenery using the received semantic information.

\begin{figure}[t]
  \centering
  \includegraphics[width=0.4\textwidth]{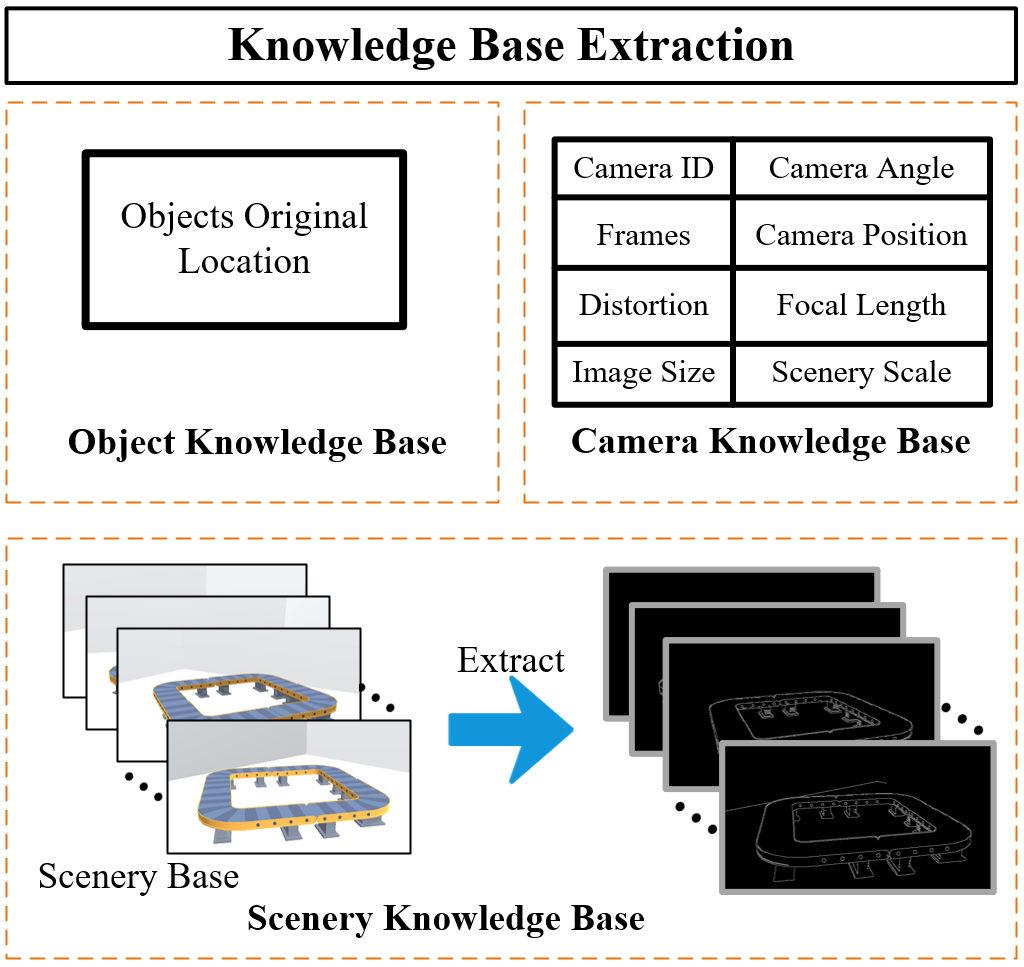} 
  \caption{Knowledge Base Extraction} 
  \label{baseKnowledge_extraction} 
\end{figure}

\subsection{Knowledge Base Extraction}
As shown in Fig. \ref{baseKnowledge_extraction}, compared to the general metaverse construction task that only relies on images input, our proposed GSC framework integrates knowledge base $\mathcal{B}$ at the receiver for metaverse rendering. 
In detail, multiple components in the metaverse, such as stationary background objects denoted as $O$, remain stable. Background objects include the background of a factory or fixed objects like conveyor belts. In contrast, movable objects $\mathcal{M}$, such as robotic arm and a moving box, are in motion.
The knowledge base refers to same information within the stable and moving components through operation. 
These knowledge base only needs to be transmitted at the beginning of the metaverse application and thus alleviate the bandwidth requirements. 
We define the knowledge base as scenery knowledge base, camera knowledge base, and object knowledge base, that is derived from the original metaverse scenery.
\begin{itemize}
    \item The scenery knowledge base includes the canny image set $\mathcal{V}_c$, extracted from the image set $\mathcal{V}_0$ of the scenery at time slot $t=0$. These canny images contain information about the metaverse's stationary background and can serve as rotation information in the image generation process.
    \item The camera knowledge base represents the UAV camera parameters $\mathcal{C}$, as described in Eq. (\ref{camera_information}), which include the UAV camera's ID, camera angle, distance, focal length, etc. These parameters will be used as input for 3D metaverse construction process based on images.
    \item The object knowledge base consists of all objects' three-dimensional location coordinates at the time $t=0$. These coordinates will be used to attach objects within metaverse scenery, allow them to be correctly placed after being constructed based on images.
\end{itemize}
\subsection{Semantic Encoder}
\begin{figure}[t]
  \centering
  \includegraphics[width=0.45\textwidth]{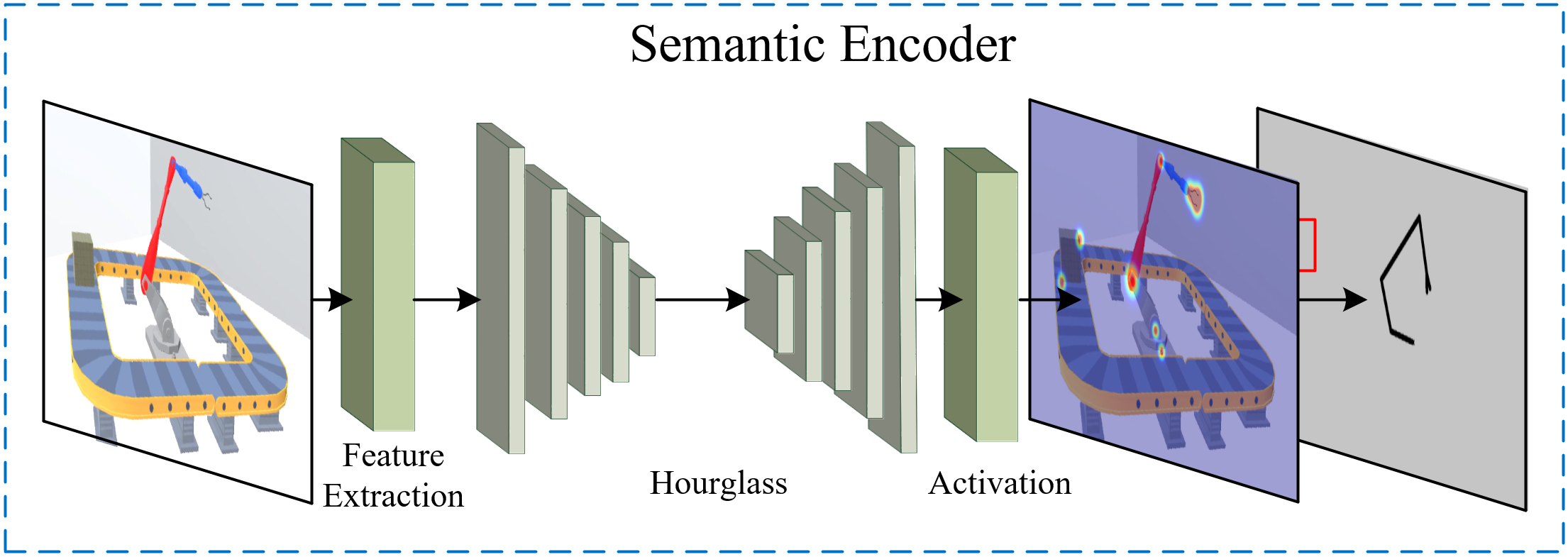} 
  \caption{Semantic Encoder} 
  \label{Semantic_encoder} 
\end{figure}

Section II.B details that the operational status of objects in the metaverse is considered essential for representing the state of the metaverse. Thus, in our industrial factory scenario, the positions and movements of moving boxes and robotic arms need to be accurately transmitted. Therefore, we define key points that precisely capture object movements and positions as semantic information.
The architecture of the semantic encoder is detailed in Fig. \ref{Semantic_encoder}, which generates nine heatmaps through a series of convolutional and deconvolutional operations. Each heatmap corresponds to the predicted location of a specific keypoint.
The coordinates of the key points $\overrightharp{K}_i$ can be extracted from the heatmaps as
\begin{equation}
\overrightharp{K}_i=\mathcal{H}(\mathbf{I}_{i}, \delta_{\text{h}})=\arg \max_{(i, j)} \text{H}_k(i, j)=(x_i,y_i),
\end{equation}
where $\delta_{\text{h}}$ denotes the neural network parameters in the HgNet, and $\overrightharp{K}_i$ represent the key points of the scenery, $\text{H}_k$ represents the heatmap of the $k$-th key point, and the location of the maximum value corresponds to the predicted coordinates of the keypoint. The loss function of HgNet is to minimize the Euclidean distance between the predicted key point and groundtruth, which is represented as
\begin{equation}
\sqrt{|\overrightharp{K}_i-\overrightharp{K}_{\text{g}}|^{2}}
= \sqrt{\frac{1}{N} \sum_{i=1}^N \left( \left(x_i - {x}_g\right)^2 + \left(y_i - {y}_g\right)^2 \right)},
\end{equation}
where $\overrightharp{K}_{g}$ represents the ground truth keypoint information related to the location and operational status of the moving box and robotic arm. 
The semantic encoder architecture and training parameters are shown in Table \ref{tab:Hgnet_arch}.

\begin{figure*}[t]
  \centering
  \includegraphics[width=\textwidth ]{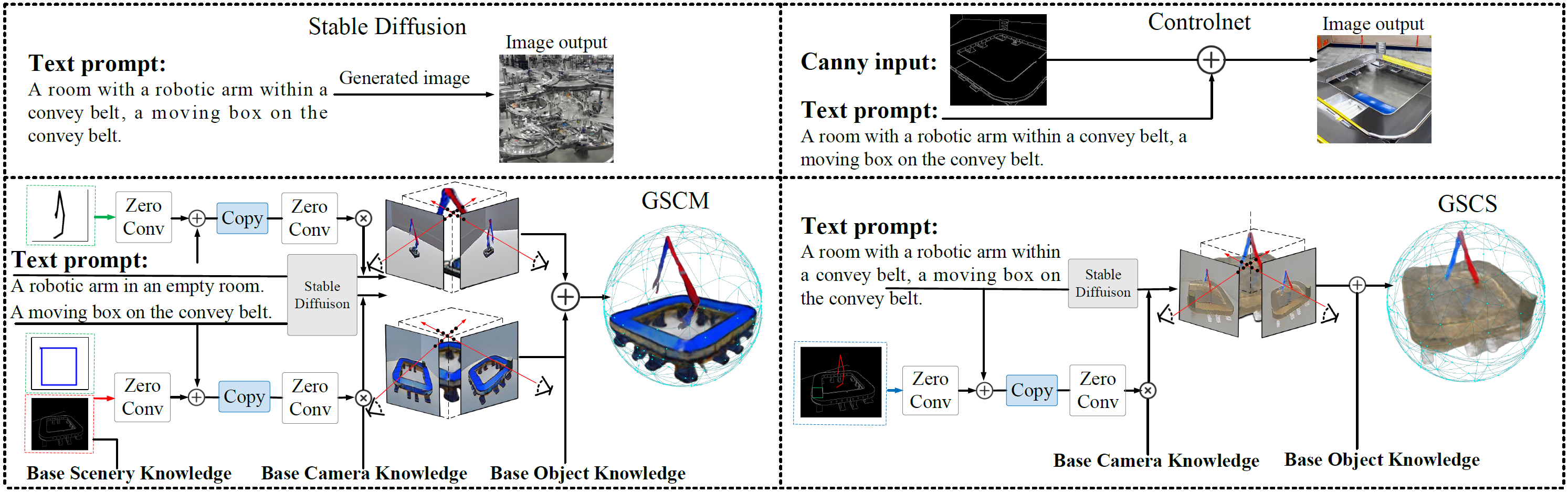} 
  \caption{Semantic Decoder for Metaverse Construction} 
  \label{metaverse_nerf} 
\end{figure*}

\begin{table}[t]
  \centering
  \caption{HgNet Architecture}
  \label{tab:Hgnet_arch}
  \begin{tabular}[l]{@{}lc}
  \toprule      
      \textbf{\emph{\textcolor{black}{HgNet Parameters}}}  & \textbf{\emph{\textcolor{black}{Input Value}}}\\
        \textcolor{black}{Residual Block 1 (up1)} & \textcolor{black}{(256,256,128)} \\ 
        \textcolor{black}{MaxPool} &  (256,256,128) \\ 
        \textcolor{black}{Recursive HourGlass (n=3)} &  (128,128,128) \\ 
        \textcolor{black}{Upsample (up2)}  & (128,128,128) \\ 
       \textcolor{black}{Final Output (up1 + up2)} & (256,256,128)\\ 
        \textcolor{black}{Batch Size} & \textcolor{black}{40} \\ 
        \textcolor{black}{Momentum} & \textcolor{black}{0.9} \\ 
        \textcolor{black}{Weight Decay} & $10^{-4}$ \\ 
    \bottomrule
  \end{tabular}
\end{table}

\subsection{Semantic Decoder}
The semantic decoder is designed to construct metaverse scenery using extracted semantic information and shared knowledge base. The decoder process includes both image generation and metaverse construction.
\subsubsection{Image Generation}

The designed semantic decoder first generates images based on the received semantic key point information. To achieve this, we develop the first module of our semantic decoder based on the SD algorithm. Specifically, SD uses a reverse diffusion process, where noise is iteratively removed from a random noise vector to produce coherent images. Mathematically, the SD model uses a series of denoising steps, represented by a sequence of latent variables, to estimate a noise-removed image that matches the desired target. The joint probability distribution for conditional image generation, incorporating various input modalities like keypoints, line drawings, and text prompts. Inspired by the implementation of SD for image generation as presented in \cite{zhang2023adding}, we design a ControlNet-based multi-rotation image generation SD algorithm. As shown in Fig. \ref{metaverse_nerf}, this algorithm incorporates additional conditional text and image prompts, such as robotic arm key points and canny edge images, into the SD model to generate images depicting multiple rotational views of the same scene, the probability distribution for generating the final image under given condition can be expressed as
\begin{equation}
p_\theta \left(\mathbf{I}^{\prime}_0 \mid t_{\text{p}}, \mathbf{I}_{\text{c}}, \overrightharp{K}^{\prime}\right) = \int p_\theta\left(\mathbf{I}^{\prime}_{0:\text{T}} \mid t_{\text{p}}, \mathbf{I}_{\text{c}}, \overrightharp{K}^{\prime}\right) d \mathbf{I}^{\prime}_{1:\text{T}} ,
\end{equation}
where $t_{\text{p}}$ represents the text prompt describing the image content and style, $\mathbf{I}_{\text{c}}$ denotes the canny edge images from the knowledge base, and $\overrightharp{K}^{\prime}$ represents the key point information provided as input, and $p_\theta(\mathbf{I}^{\prime}_{0:\text{T}} \mid t_{\text{p}}, \mathbf{I}_{\text{c}}, \overrightharp{K}^{\prime})$ indicates the joint probability of generating an image conditioned on these inputs through the diffusion process, starting from the initial step $\text{T}$.
With these control conditions, the reverse diffusion process becomes a conditional generation process that ensures the generated image aligns with both the text prompt and image control inputs. At each stage of the generation process, the generated image is assumed to follow a Gaussian distribution, described as
\begin{equation}
\mathbf{I}^{\prime}_{\text{T}-1} \sim \mathcal{N}\left(\mu_\theta\left(\mathbf{I}^{\prime}_\text{T}, \text{T}, \mathbf{t}_p, \overrightharp{K}^{\prime}, \mathbf{I}_{\text{c}}\right), \Sigma_\theta\left(\mathbf{I}^{\prime}_\text{T}, \text{T}\right)\right),
\end{equation} 
where $\mu_\theta(\cdot)$ represents the mean function that determines the most likely generated image $\mathbf{I}^{\prime}_{\text{T}-1}$, and $\Sigma_\theta(\cdot)$ denotes the covariance matrix that governs the uncertainty in the generation process.

\subsubsection{Metaverse Construction}
In the second step, the designed semantic decoder constructs the metaverse scenery using the images generated in the first step. To do so, we develop the second module of our semantic decoder based on NeRF, a neural network-based method for synthesizing 3D views from images by estimating the density and radiance at sampled points within the metaverse.
The algorithm learns a mapping from spatial coordinates and viewing directions to color and density values, enabling high-fidelity 3D reconstruction.
A NeRF algorithm is utilized to reconstruct the scenery by learning the volumetric density and color values at each point $\mathbf{r}(l)$ of the input image, the point $\mathbf{r}(l)$ is represented as
\begin{equation}
\mathbf{r}(l)=\overrightharp{o}+l \cdot \overrightharp{d},
\end{equation}
where $\overrightharp{o}$ represents the UAV’s position, $l$ is the distance between the pixel coordinate and the UAV’s position, and $\overrightharp{d}$ is the ray direction computed from pixel coordinate. The ray direction $\overrightharp{d}$ is represented as
\begin{equation}
\begin{aligned}
&\vec{d}=\frac{1}{\sqrt{u^{\prime 2}+v^{\prime 2}+1}}\left(\begin{array}{ccc} u',  v',  1 \end{array}\right)^\text{T}\\
 \quad &u^{\prime}=\frac{u-c_x}{f_x}, \quad v^{\prime}=\frac{v-c_y}{f_y},
\end{aligned}
\end{equation}
$u$ and $v$ denote the pixel coordinates of the input image, $c_x$ and $c_y$ represent the principal point offsets, and $f_x$ and $f_y$ are the focal lengths in the $x$ and $y$ directions, which are the fixed camera parameters described in Eq. (\ref{camera_information}).
The NeRF algorithm synthesizes realistic novel views by predicting the color and density along rays emitted from the UAV camera. The predicted image $\overline{\mathbf{I}}_t^{\prime}$ from the rendered metaverse scenery is expressed as
\begin{equation}
\label{NeRF_algo}
\begin{gathered}
\overline{\mathbf{I}}_t^{\prime} = \int_{l_s}^{l_e} T(l) \sigma(\mathbf{r}(l)) \mathbf{c}(\mathbf{r}(l), \overrightharp{d}) \, dl,
\end{gathered}
\end{equation}
where the accumulated transmittance $T(l)$ quantifies the likelihood that a ray travels without being occluded, and $\sigma(\mathbf{r}(l))$ denotes the scene’s density at point $\mathbf{r}(l)$. 
\subsection{Framework Design}
To better clarify the differences and advantages between semantic communication and conventional metaverse construction and effectively validate the performance of our proposed GSC framework, we design various types of transmitted semantic information and knowledge base, as shown in Fig. \ref{metaverse_nerf}. These include model-based GSC (GSCM) and scenery-based GSC (GSCS), to provide a comparative foundation for evaluation.

\textbf{GSCS}: 
The GSCS framework generates multi-angle images of the metaverse scenery by utilizing key points, moving box locations, and canny images together. The image generation process is described as
\begin{equation}
\label{sd2}
\hat{\mathbf{I}}_t^{\prime}=\mathcal{D}\left(t_p, \mathbf{I}_\text{c}, \vec{K}_t^{\prime}, \vec{B}_t^{\prime}, \omega_c, \omega_k, \omega_b, \theta\right),
\end{equation}
where \(\omega_{c}\), \(\omega_{k}\), and \(\omega_{b}\) represent the control weights for the canny image, robotic arm, and moving box. These weights determine the clearance of each component in the generated image $\hat{\mathbf{I}}_t^{\prime}$. 

\textbf{GSCM}: 
The GSCM framework generates metaverse scenery images in two steps: 1) Using key points as input to generate images of the robotic arm. 2) Using moving box information and canny images to generate images of the metaverse's stationary scenery. In this framework, the image generation process is described as
\begin{equation}
\label{sd1}
\hat{\mathbf{I}}_t^{\prime}= \begin{cases}\mathcal{D}\left(t_{p}, \mathbf{I}_\text{c}, \vec{B}_t^{\prime}, \theta_{i}\right), & \text { if } i=1 \\ \mathcal{D}\left(t_{p}, \vec{K}_t^{\prime}, \theta_{i}\right), & \text { if } i=2\end{cases},
\end{equation}
where $t_{p}$ represents the text prompt defined either to generate images of the robotic arm or the stationary scenery.

\section{Optimal Transport-enabled Denoiser}
\begin{figure}[t]
  \centering
  \includegraphics[width=0.45\textwidth ]{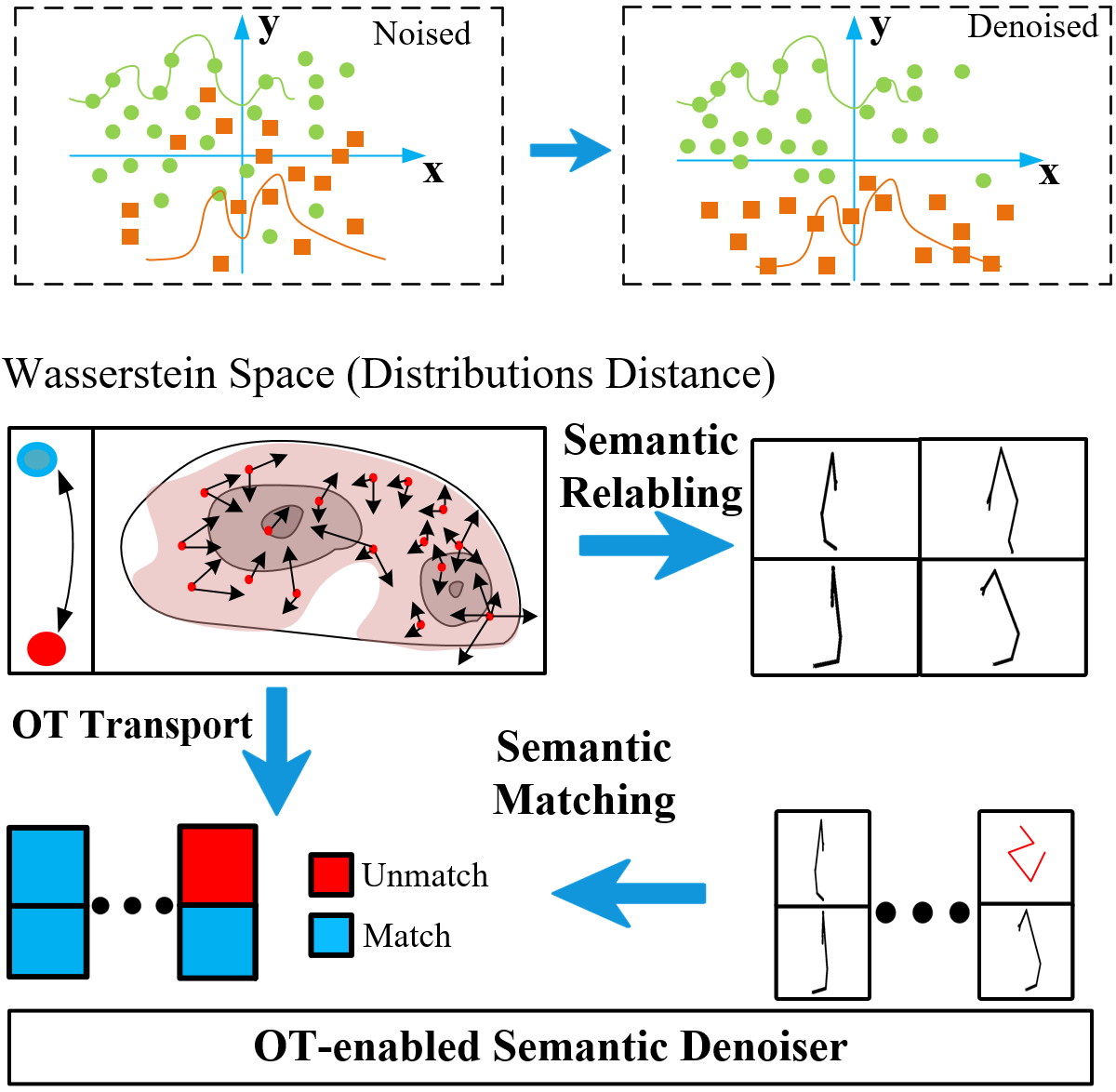} 
  \caption{Semantic Denoiser} 
  \label{denoiser_image} 
\end{figure}

In this section, we present the design principle and algorithm of our semantic OT-based denoiser, as shown in Fig. \ref{denoiser_image}. 
At the receiver, we propose a relaxed OT optimization to compute the transport matrix between the transmitter and receiver. The  OT optimization selectively corrects distribution shifts caused by wireless channel fading and noise and then used the corrected information for later semantic decoder to reconstruct the metaverse.

\subsection{Relaxed Optimal Transport Optimization}
Inspired by the OT algorithm on large volume data optimization \cite{sana2023semantic}, we design an OT-enabled semantic denoising algorithm involves relaxing the row and column constraints of the received data individually instead of enforcing both simultaneously for semantic communication in metaverse construction.
Given the received key point vectors $\overrightharp{K}^{\prime}=\left\{(x^{\prime}_i,y^{\prime}_i)\right\}_{i=1}^n$ at the receiver side and the original point vectors at the transmitter side $\overrightharp{K}=\left\{(x_j,y_j)\right\}_{j=1}^n$, with their respective probability distributions $D_{r}$ and $D_{t}$, the goal of the OT-enabled semantic denoising algorithm is to find an optimal transport matrix ${T}_{ij}$ that minimizes the correction cost from $D_{r}$ to $D_{t}$, which can be represented as
\begin{equation}
\begin{aligned}
\label{OT_objective}
\min _T \sum_{i=1}^n \sum_{j=1}^n T_{i j} c_{i j} &+ \eta \sum_{i=1}^n \sum_{j=1}^n T_{i j}\left(\log T_{i j} - 1\right), \\
\text{subject to} &\quad \sum_{j=1}^n T_{i j} = p_i,\\
&\quad \sum_{i=1}^n T_{i j} = q_i,
\end{aligned}
\end{equation}
where $p_i$ represents the row marginal distribution of the received data. $T_{i j}$ represents the amount of mass transported from point set $\overrightharp{K}^{\prime}$ to point set $\overrightharp{K}$, and $c_{ij}$ is the transportation cost between point $(x^{\prime}_i,y^{\prime}_i)$ and $(x_j,y_j)$. 
This cost represents the complexity of redistributing one set of vectors to match the other and is calculated using the Euclidean distance between the sample points, and can be expressed as
\begin{equation}
C_{i j}=\text{dist}\left((x^{\prime}_i, y^{\prime}_i),(x_j, y_j)\right),
\end{equation}
where the $\text{dist}(\cdot)$ is the Euclidean distance calculated from two points. 
The goal of the OT is to ensure that the probability distributions of the key point vectors at the receiver and sender sides are matched, while simultaneously minimizing the transportation cost. To address this, the Lagrange function is introduced to incorporate the constraints in Eq. (\ref{OT_objective}) into the objective function using the Lagrange multiplier method, the problem is transformed into an unconstrained optimization task, which can be expressed as
\begin{equation}
\begin{aligned}
L = \sum_{i=1}^n \sum_{j=1}^n T_{i j} c_{i j} 
&+ \eta \sum_{i=1}^n \sum_{j=1}^n T_{i j} \left( \log T_{i j} - 1 \right), \\
&- \sum_{i=1}^n \lambda_i \left( \sum_{j=1}^n T_{i j} - p_i \right),
\end{aligned}
\end{equation}
where $\lambda_i$ is the Lagrange multiplier, taking the partial derivative of $L$ with respect to $T_{i j}$ and setting it to zero is done to find the minimum value of the function. This step helps determine an expression for $T_{i j}$ that satisfies the optimization conditions, which can be expressed as
\begin{equation}
\begin{aligned}
\label{row_matrix_results}
\frac{\partial L}{\partial T_{i j}}=c_{i j}+\eta\left(\log T_{i j}-1\right)-\lambda_i=0
\end{aligned}
\end{equation}
\begin{equation}
\text{where} \ T_{i j}=e^{\frac{\lambda_i-c_{i j}}{\eta}},   
\end{equation}
Then, the OT problem involves aggregating over $j$ and applying the row constraint $\sum_{j=1}^n T_{i j}=p_i$. By incorporating the marginal distribution constraint, the optimal transport matrix that satisfies the conditions Eq. (\ref{OT_objective}) can be determined, and it can be integrated with Eq. (\ref{row_matrix_results}), expressed as
\begin{equation}
\begin{aligned}
\label{lamda_results}
p_i&=\sum_{j=1}^n e^{\frac{\lambda_i-c_{i j}}{\eta}},\\
\end{aligned}
\end{equation}
\begin{equation}
\text{where} \ \lambda_i =\eta \log \left(\frac{p_i}{\sum_{j=1}^n e^{-\frac{c_j}{\eta}}}\right).   
\end{equation}
Thus, the relaxed transport matrix $\textbf{T}_U$ under the row constraint is calculated by substituting Eq. (\ref{lamda_results}) into Eq. (\ref{row_matrix_results}), which is represented as
\begin{equation}
\textbf{T}_U=\tilde{C} \cdot \operatorname{diag}\left(\frac{p}{\mathbf{1}_n \tilde{C}}\right),
\end{equation}
where $\tilde{C}=e^{-\frac{C}{\eta}}$, $\mathbf{1}_n \tilde{C}$ refers to the matrix product of this vector of ones with the matrix, and ${p}$ represents the uniform importance weight of each point in the source set
Similarly, as for the column constraint, we consider the relaxation of the column constraint, where the transport matrix $T$  is optimized with only the column sums constrained $\text { subject to } \quad \sum_{i=1}^n T_{i j}=q_j$.
Following a similar procedure as above, the relaxed transport matrix $T_V$ under the column constraint is
\begin{equation}
\textbf{T}_V=\operatorname{diag}\left(\frac{q}{\mathbf{1}_n \tilde{C}^T}\right) \cdot \tilde{C}.
\end{equation}
To approximate the solution under the original full constraint, we combine the solutions from the relaxed row and column constraints and use the element-wise maximum of the matrices $\textbf{T}_U$ and $\textbf{T}_V$
\begin{equation}
\begin{aligned}
\textbf{T}^*&=\max \left(\textbf{T}_U, \textbf{T}_V\right),\\
\textbf{T}^*&=\max \left(\tilde{C} \cdot \operatorname{diag}\left(\frac{p}{\mathbf{1}_n \tilde{C}}\right), \operatorname{diag}\left(\frac{q}{\mathbf{1}_n \tilde{C}^T}\right) \cdot \tilde{C}\right).
\end{aligned}
\end{equation}
The OT-enabled semantic denoising algorithm implementation achieves a computational complexity of $O\left(n^2\right)$, compared to the traditional OT problem, which uses the Sinkhorn-Knopp algorithm with $O\left(n^3\right)$ complexity, making it more efficient for large-scale problems.

\begin{algorithm}[t]
\caption{Relaxed OT Denoising Algorithm}
\label{Unified_OT_Algo}
\begin{algorithmic}[1]
\State \textbf{Input:} Transport cost $c_{ij}$, regularization parameter $\epsilon$, maximum iterations $N_{\text{max}}$, Image set $\mathbf{I}$ with corresponding captured angles $\theta_i$, key points set $\mathbf{K}_i$, threshold $\delta$
\State \textbf{Output:} Denoised key points $\mathbf{\hat{K}}_i$

\State Initialize the transport cost matrix $\tilde{C}$ as:
$\tilde{C}_{ij} = \exp\left(-\frac{c_{ij}}{\epsilon}\right)$
\State Normalize $\tilde{C}$:
$\tilde{C} \leftarrow \frac{\tilde{C}}{\sum_{i,j} \tilde{C}_{ij}}$
\State Initialize marginal distributions $p_i$ and $q_j$ from $\mathbf{K}_i$.

\For{each image $\textbf{I}_i$}
    \State Initialize vectors $u$ and $v$ to all ones for row and column updates, respectively.
    
    \State \textbf{Step 1: Row Constraint Optimization (\(\textbf{T}_U\))}:
    \For{$n = 1$ to $N_{\text{max}}$}
        \State Update $u$ as: $u \leftarrow \frac{p}{\tilde{C} v}$
        \State Normalize $\textbf{T}_U$ using $u$: $\textbf{T}_U = \text{diag}(u) \cdot \tilde{C}$
    \EndFor

    \State \textbf{Step 2: Column Constraint Optimization (\(\textbf{T}_V\))}:
    \For{$n = 1$ to $N_{\text{max}}$}
        \State Update $v$ as: $v \leftarrow \frac{q}{\tilde{C}^\top u}$
        \State Normalize $\textbf{T}_V$ using $v$: $\textbf{T}_V = \tilde{C} \cdot \text{diag}(v)$
    \EndFor

    \State \textbf{Step 3: Combining Relaxed Solutions (\(\textbf{T}^*\))}:
    \State Final transport matrix $\textbf{T}^*= \max(\textbf{T}_U, \textbf{T}_V)$ 
    \State \textbf{Step 4: Key Point Filtering and Update:}
    \For{each image pair $(\textbf{I}_{i-n}, \textbf{I}_{i+n})$}
        \If{$\left|\mathbf{K}_i - \frac{\mathbf{K}_{i-n} + \mathbf{K}_{i+n}}{2} \right| \leq \delta$}
            \State ${F}_{i} \gets 1$
        \EndIf
    \EndFor

    \State Update key points for unfiltered data:
    $\mathbf{\hat{K}}_{i} \gets (\textbf{T}^* \cdot \mathbf{K}_i) \quad \text{where } ({F}_{i} == 0)$
\EndFor
\State \textbf{Return} Denoised key points $\mathbf{\hat{K}}_i$
\end{algorithmic}
\end{algorithm}

\begin{figure*}[t]
  \centering
  \includegraphics[width=\textwidth]{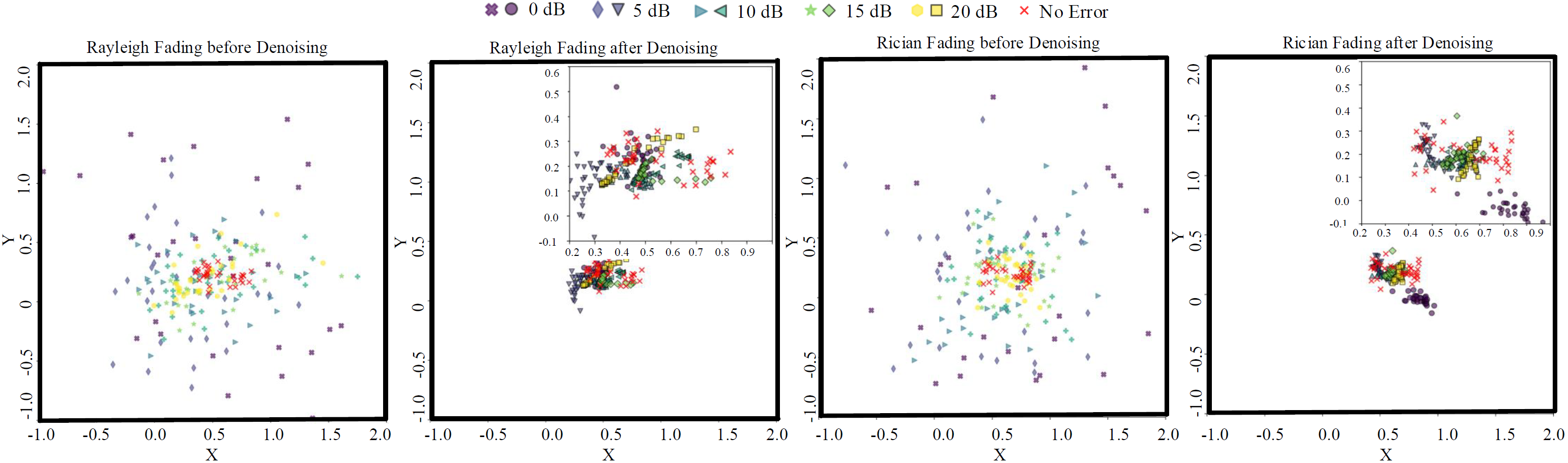} 
  \caption{Denoising Performance under Rayleigh and Rician Channels} 
  \label{OT_denoise} 
\end{figure*}

\subsection{Semantic Selective Correction}
Under the effect of wireless channel, we noticed that not all point vectors were severely blurred after wireless communication. 
Some point vectors, especially under high SNR conditions, were less affected by the wireless communication channel and, therefore, did not require denoising.
Therefore, we implement a semantic selective correction to detect potential transmission errors in the data from the received metaverse semantic information. 
The main reason is to check the correspondence of key points across images from different angles after wireless communication, as demonstrated in the \textbf{Algorithm} \ref{Unified_OT_Algo}. 
Ideally, when the rotational angles are uniform, the key point displacement should proportionally increase with the angle, and thus significant deviations indicate potential noise or transmission errors.
For key points in the $(i-n)$-th and $(i+n)$-th views, the selective filter flag $F_{i}$ is determined based on the Euclidean distance between them, calculated as
\begin{equation}
F_{i} =
\begin{cases} 
1, & \text{if } \left|\overrightharp{K}_i - \frac{\overrightharp{K}_{i-n} + \overrightharp{K}_{i+n}}{2} \right| > \delta, \\ 
0, & \text{if } \left|\overrightharp{K}_i - \frac{\overrightharp{K}_{i-n} + \overrightharp{K}_{i+n}}{2} \right| \leq \delta.
\end{cases}    
\end{equation}
If the correspondence between key points from different angles does not align, the correction and labeling process is used to precisely align these points with OT to reduce wireless communication noise.

We define different OT denoising semantic communication frameworks based on the proposed GSCM and GSCS frameworks, including:
(1) GSCS-OT: Based on \textbf{Algorithm} 1, selective semantic correction was first performed under different SNR conditions. The corrected semantic information was then incorporated into the GSCS framework, along with the image generation process described in Eq. (\ref{sd2}).
(2) GSCM-OT: Selective semantic correction was first performed under different SNR conditions using \textbf{Algorithm} 1. The corrected semantic information was then incorporated into the GSCM framework, along with the image generation process described in Eq. (\ref{sd2}), to generate various metaverse objects.

\section{Experiment Evaluation}
In this section, we evaluate the performance of our proposed GSC framework and compare it with the traditional metaverse construction framework discussed in Sections II and III. The results of our proposed GSC framework are presented as follows. Section IV-A provides insights into the configuration of metaverse scenarios and OT denoiser, while Section IV-B presents the experimental results on the semantic information extraction accuracy achieved by HgNet. 
Additionally, we assess various metrics described in III-C for our proposed GSC framework including KPE, P2Point, and latency.

\begin{table}[t]
  \centering
  \caption{Experiment Setup}
  \label{tab:Experiment_Setup}
  \begin{tabular}[l]{@{}lc}
  \toprule      
      \textbf{\emph{Metaverse Simulation}}  & \textbf{\emph{Value}}\\
        FPS &  60 Hz \\ 
        Robotic arm point number &  7 \\ 
        Moving box number&  2 \\ 
        UAV number&  36 \\ 
       Image resolution &  1200 x 600 \\ 
      \textcolor{black}{Channel response} &  \textcolor{black}{AWGN, Rayleigh, Rician} \\ 
             \textcolor{black}{Geometry movement range (x,y,z)} &  \textcolor{black}{([0,4], [0,4], [0,4])}\\ 
    \bottomrule
  \end{tabular}
\end{table}

\begin{figure*}[t]
\centering
\subfigure[AWGN.]
{
\includegraphics[width=0.3\textwidth]{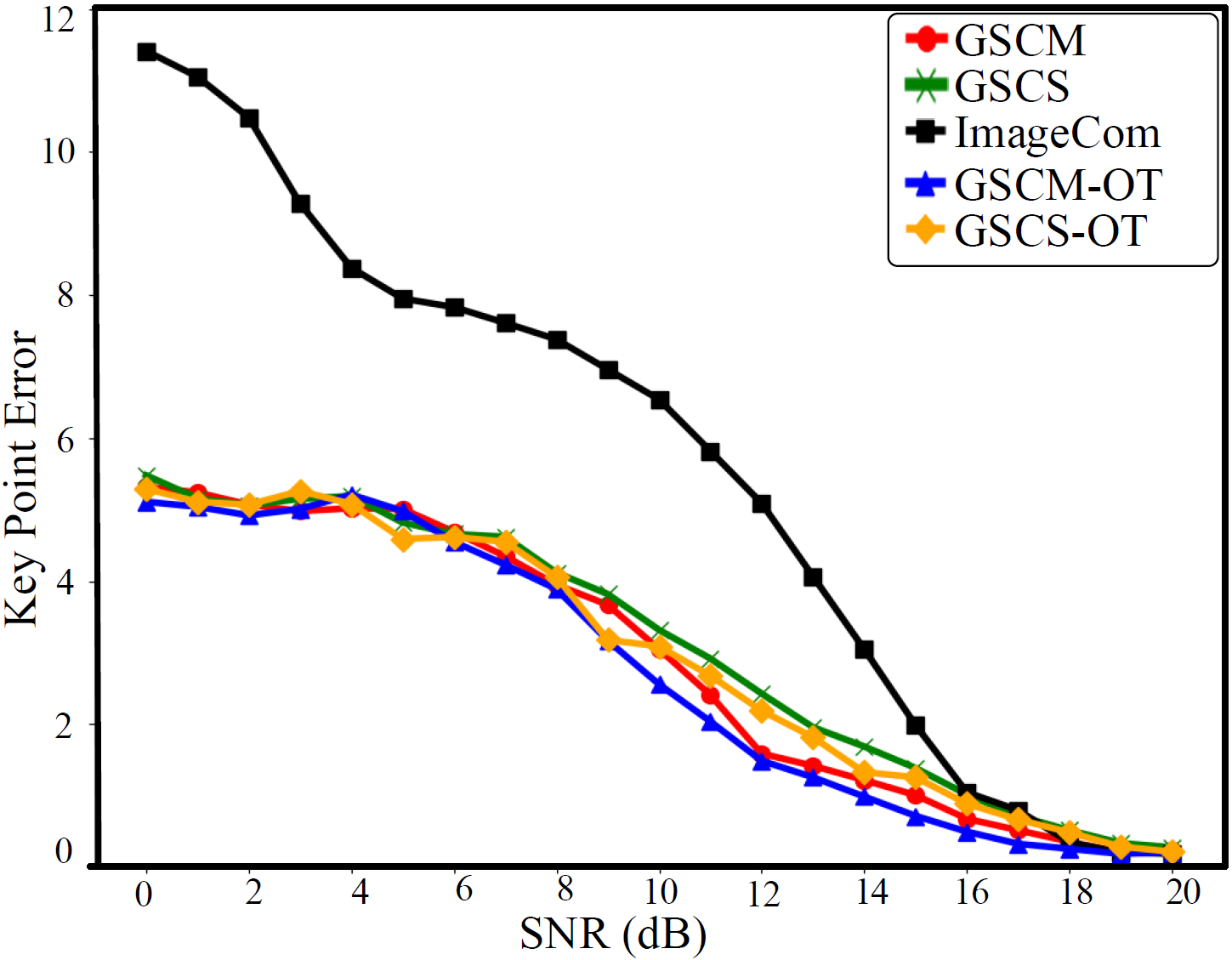}
}
\subfigure[Rayleigh.]{
\includegraphics[width=0.3\textwidth]{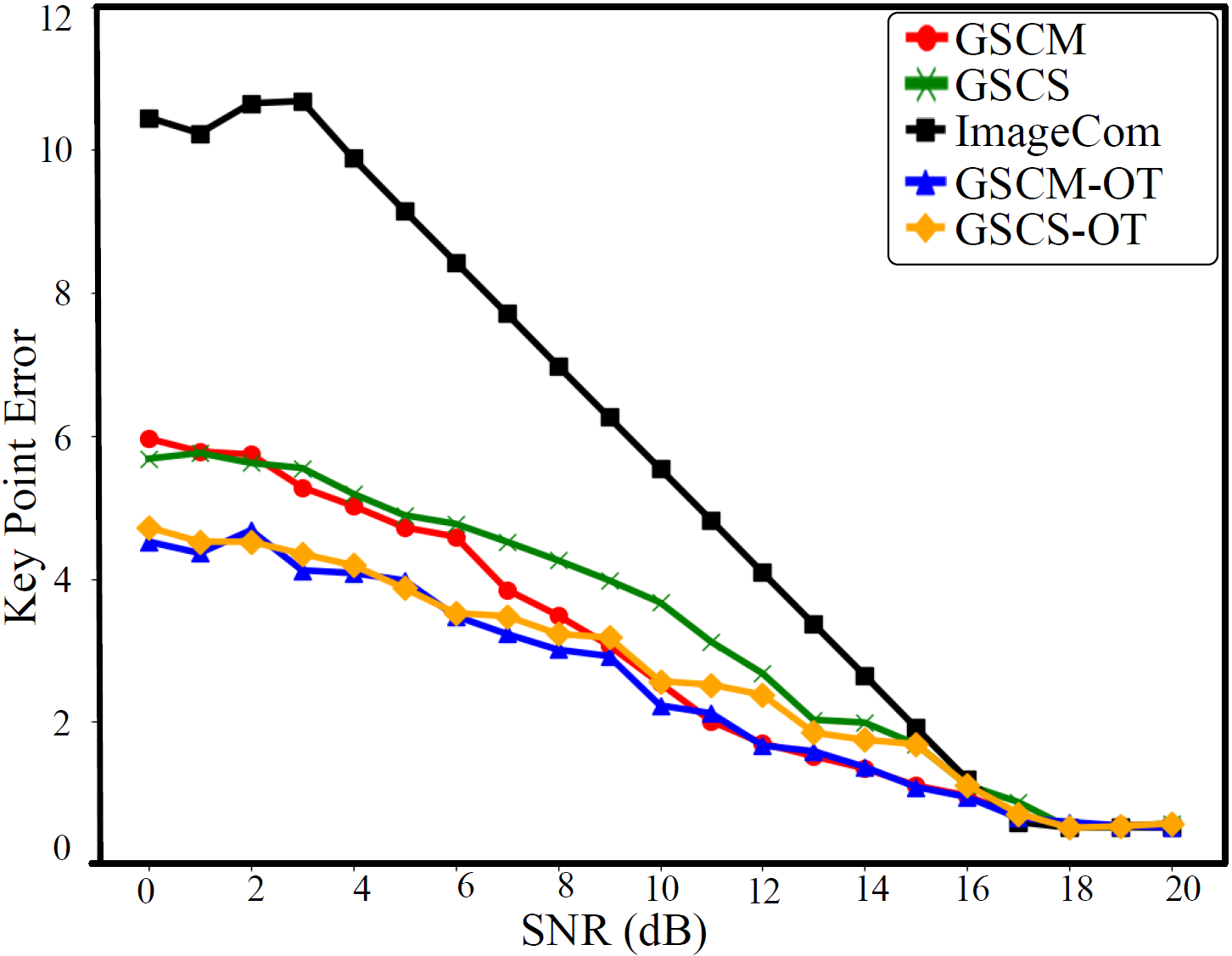}
}
\subfigure[Rician.]{
\includegraphics[width=0.3\textwidth]{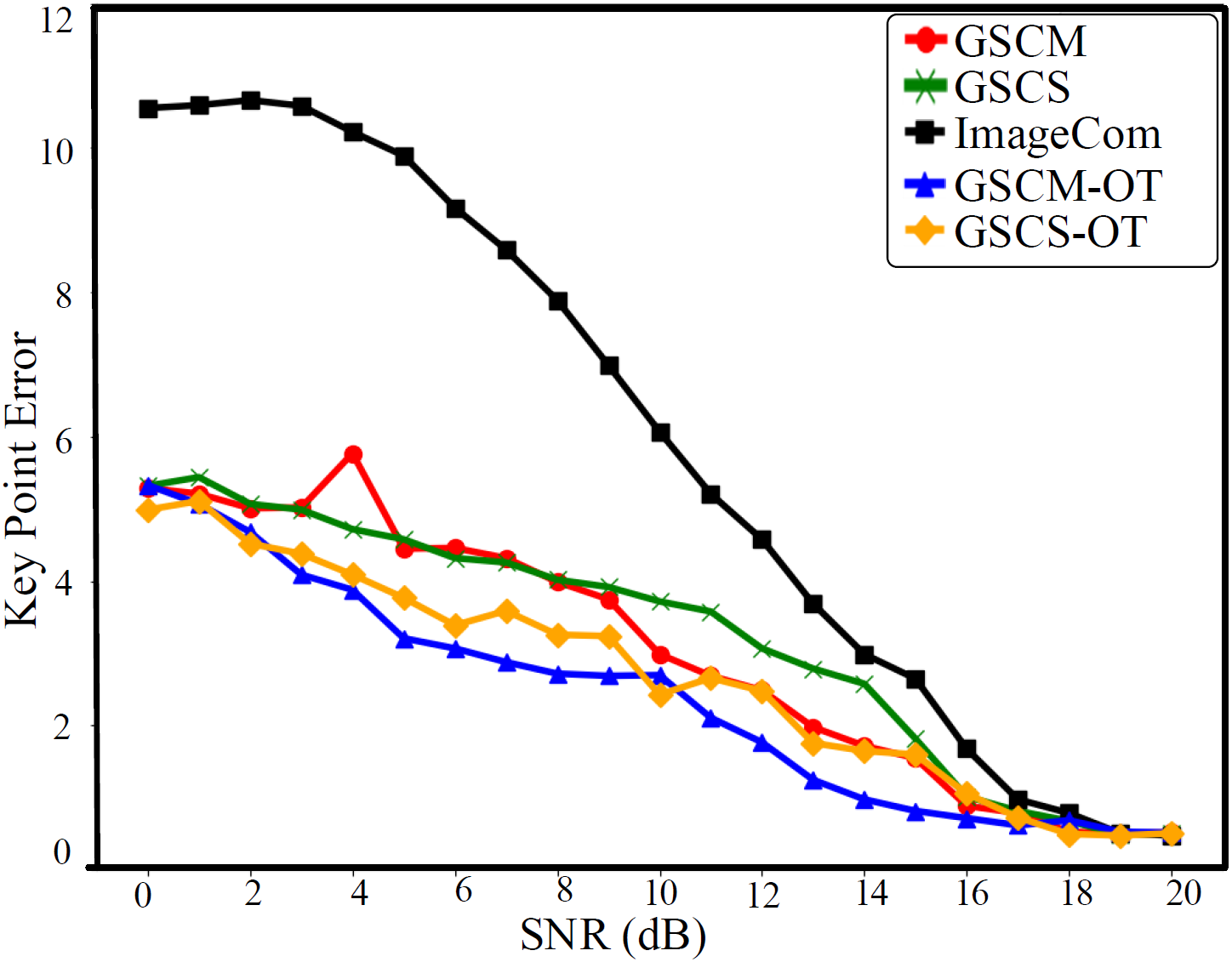}
}
\caption{Key Point Error}
\label{exkpe}
\end{figure*}

\subsection{Experiments Setup}
As for the experimental setup, we configured the wireless communication  and metaverse scenario settings as detailed in Table \ref{tab:Experiment_Setup}. These parameters include the number of captured images, channel fading type, and etc. The experimental hardware consists of Python 3.11.0 running on an Ubuntu 18.04 system, equipped with two RTX 3090 GPUs, PyTorch 2.1, and the Unity platform. Specifically, to validate the effectiveness of the OT-enabled denoiser, we introduce two additional channel models to simulate noise and fading effects in transmission, including a real additive white Gaussian noise (AWGN) channel and a Rayleigh fading channel. Both wireless channels are essential and common components in modern wireless communication systems.


\subsection{OT Denoising Capability}
\begin{figure*}[t]
\centering
\subfigure[AWGN.]
{
\includegraphics[width=0.3\textwidth]{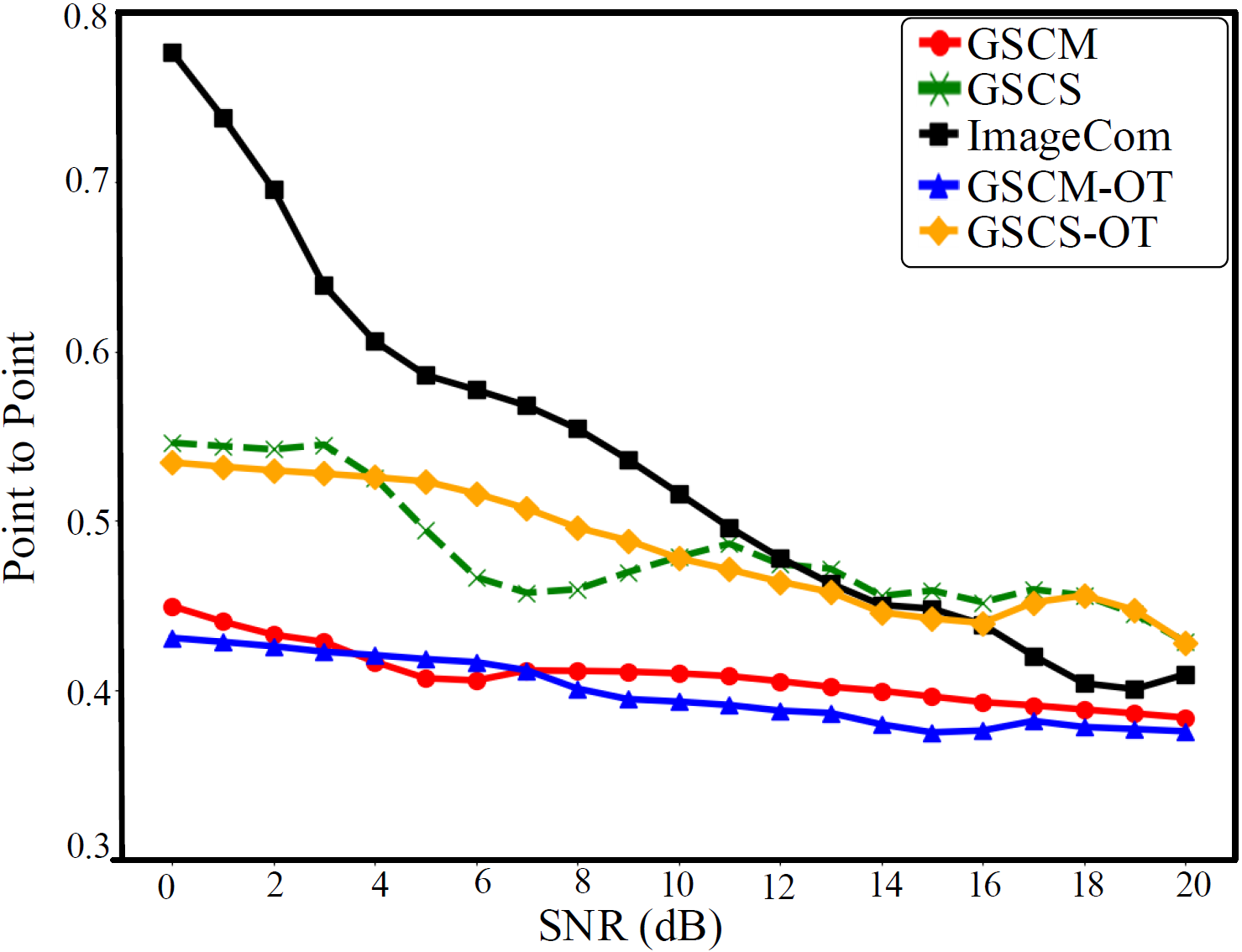}
}
\subfigure[Rayleigh.]{
\includegraphics[width=0.3\textwidth]{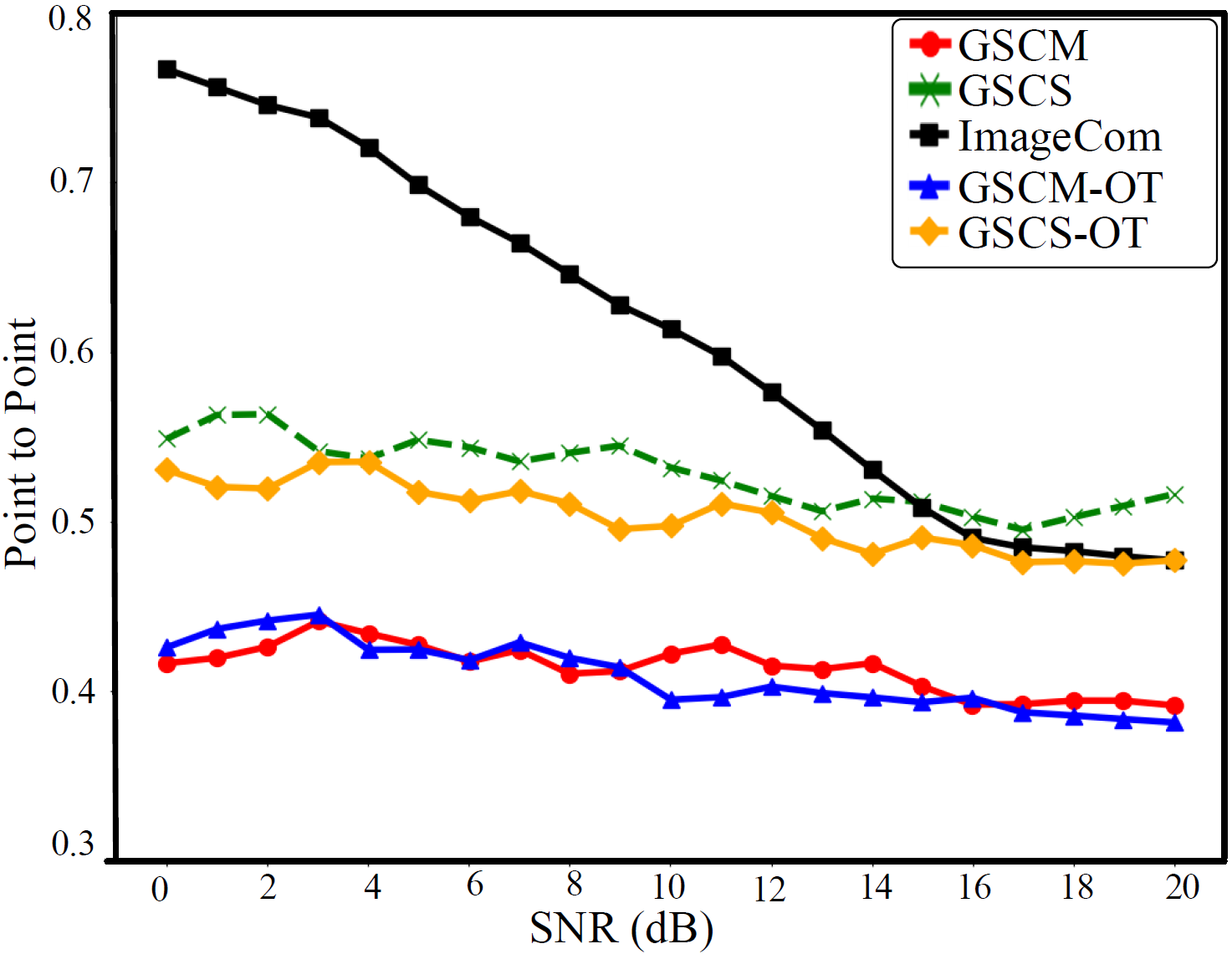}
}
\subfigure[Rician.]{
\includegraphics[width=0.302\textwidth]{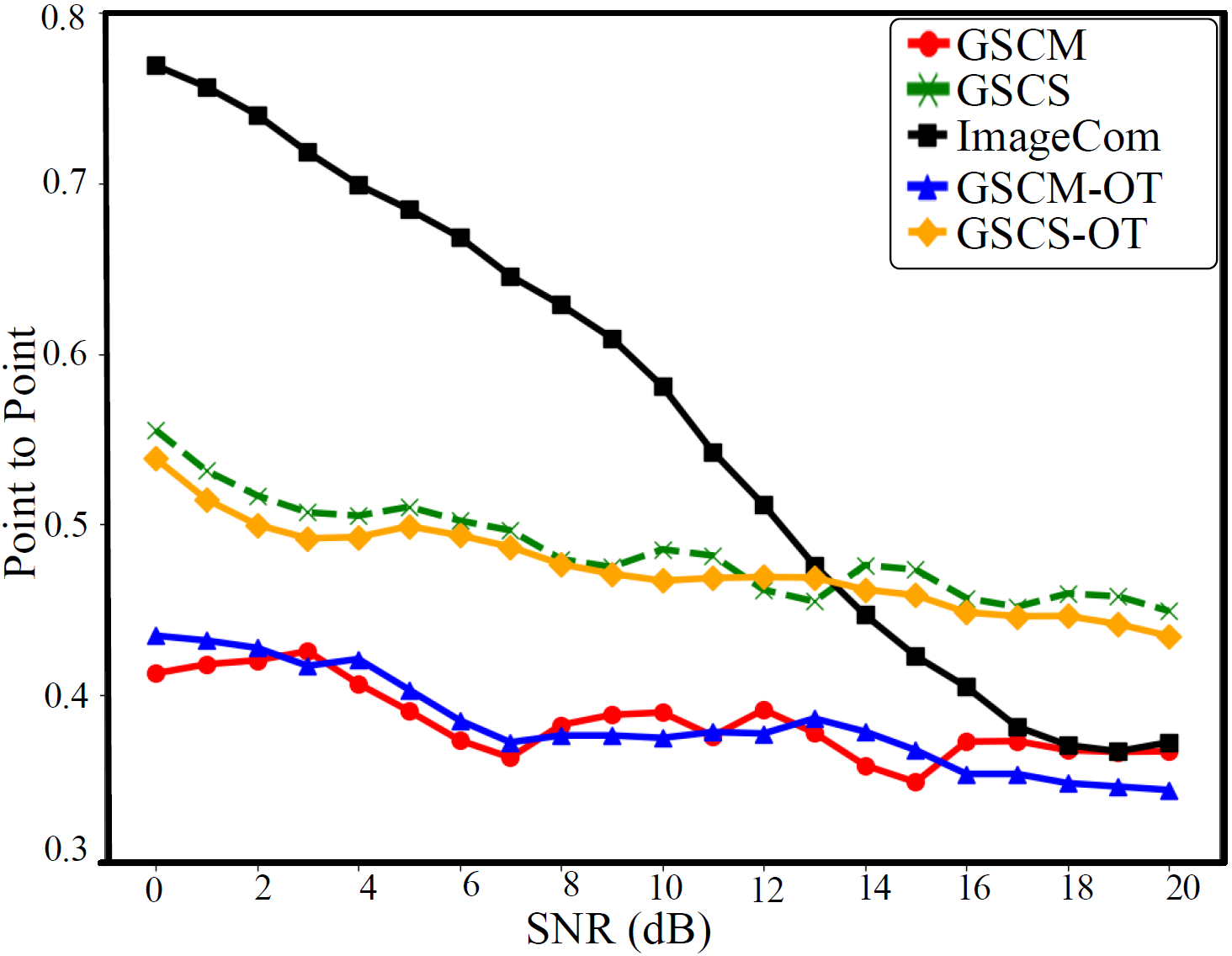}
}
\caption{Point-to-point}
\label{exp2point}
\end{figure*}

Fig. \ref{OT_denoise} plots the points difference of the OT-enabled denoiser under different wireless channel conditions. 
The point deviations increase as the SNR decreases, which occurs because lower SNR leads to higher noise levels in the wireless communication, thus causing greater point displacement.
For a better demonstration, we implement the OT-enabled denoiser varies across different wireless channels, with the performance Rayleigh similar as Rician channel.
Specifically, in the Rician channel, as the SNR decreases, the OT-enabled denoising algorithm effectively recovers the noisy points back to their original distribution. Additionally, as the SNR increases, noise gradually decreases, resulting in less variation in the received point vectors.
On the other hand, the results for the Rayleigh wireless channels show unstable denoising performance under different SNR conditions. 
When the SNR drops below 10 dB, although the denoising algorithm shows some success in point recovery, noise persists around the ground truth points, especially under 0 dB conditions, highlighting certain limitations of the OT-denoising method in a Rayleigh fading channel without a line-of-sight signal, compared with a Rician fading channel. This explains why a selective correction algorithm is first used to filter noisy points, followed by the OT-denoising algorithm for points that cannot be matched.
This phenomenon is also reflected in the P2Point error results shown in Fig. \ref{exp2point}, where the GSC framework exhibits unstable performance in P2Point and key point error as the SNR decreases in the Rician channel. While the OT-enabled denoising algorithm performs effectively in wireless channel denoising, deviations may occur under specific channel conditions and thus highlights the importance of selective correction.

\subsection{Key Points Transmission Accuracy}
Fig. \ref{exkpe} plots the KPE between the transmitter and receiver when employing our proposed GSC framework in comparison to the conventional metaverse construction framework (ImageCom) across three different wireless channel conditions. 
The error is measured using the Euclidean distance between the key points' two-dimensional positions, which is an approach for evaluating metaverse objects' status update accuracy in communication systems. 
A lower key point error implies a more accurate representation of the object running status in the metaverse, such as robotic arms and boxes. 
To explain it in more detail, Fig. \ref{exkpe} (c) plots the KPE performance under Rician channel conditions ranked as GSCM-OT $\textless$ GSCM $\textless$ GSCS-OT $\textless$ GSCS $\textless$ ImageCom suggests that the incorporation of the GSC framework, particularly with OT denoiser improves the accuracy of the transmitted key points.

With the SNR increase, the key point error shows a noticeable reduction across all frameworks. This demonstrates the general trend that higher SNR values, which indicate a cleaner communication channel, lead to more reliable and accurate transmissions. When the SNR reaches 20 dB, the key point errors for all the frameworks drop to levels indicating nearly lossless transmission. However, at 0 dB SNR condition, the differences between the GSC and ImageCom become more pronounced. The GSCS-OT framework demonstrates the best performance in terms of key point error, outperforming ImageCom by a significant decrease of 45.6\%. 
This performance improvement can be attributed to the fact that GSC, with stable diffusion and OT denoiser techniques, despite slight variations in key point positions, can still generate images of metaverse objects that are recognizable and usable for monitoring and controlling operations. 
In contrast, the ImageCom framework, which relies solely on image transmission, struggles to maintain color accuracy and often produces blurred images in low SNR and high noise environments. Under such conditions, the transmitted images become so distorted that the robotic arm and boxes are unrecognizable, resulting in a complete loss of essential objects.
Moreover, under 0 dB condition, a comparison between the frameworks with and without OT deoiser reveals the additional benefit provided by OT. The GSCM-OT framework outperforms GSCM, reducing the key point error by approximately 10.2\%, while GSCS-OT shows a similar improvement of around 5.3\% over GSCS. These reductions highlight the efficacy of OT in denoising the transmitted data, especially under AWGN channel conditions. OT helps to maintain the accuracy of key point vector transmissions by mitigating the effects of noise, which is particularly beneficial in wireless communication environments.

\subsection{Metaverse Construction Reliability}
Fig. \ref{exp2point} plots the P2Point results of different frameworks in constructing metaverse scenery at the receiver. The P2Point metric, which measures the geometric differences in 3D scenery between the transmitter and receiver, serves as an essential indicator of construction clarity and stability. A lower P2Point value indicates a more precise reconstruction of the metaverse scenery, demonstrating an improved viewing experience of the metaverse scenery.
In detail, as plotted in Fig. \ref{exp2point}(c), under the Rayleigh channel, the P2Point results are ranked as GSCM-OT $\textless$ GSCS-OT $\textless$ GSCM $\textless$ GSCS $\textless$ ImageCom. Specifically, at 0 dB conditions, GSCM-OT and GSCS-OT demonstrate improvements of 44.7\% and 29.5\%, respectively, over the ImageCom. This improvement may be attributed to the design of the GSC framework, which extracts and updates the knowledge base to maintain stability and reduce random noise in the static scenery construction process, even in scenarios where certain objects, such as the robotic arm, are positioned less precisely. 
In contrast, the ImageCom framework, which lacks the knowledge base provided for the metaverse construction, struggles to preserve structural integrity under challenging conditions.

\begin{figure}[t]
  \centering
  \includegraphics[width=0.42\textwidth]{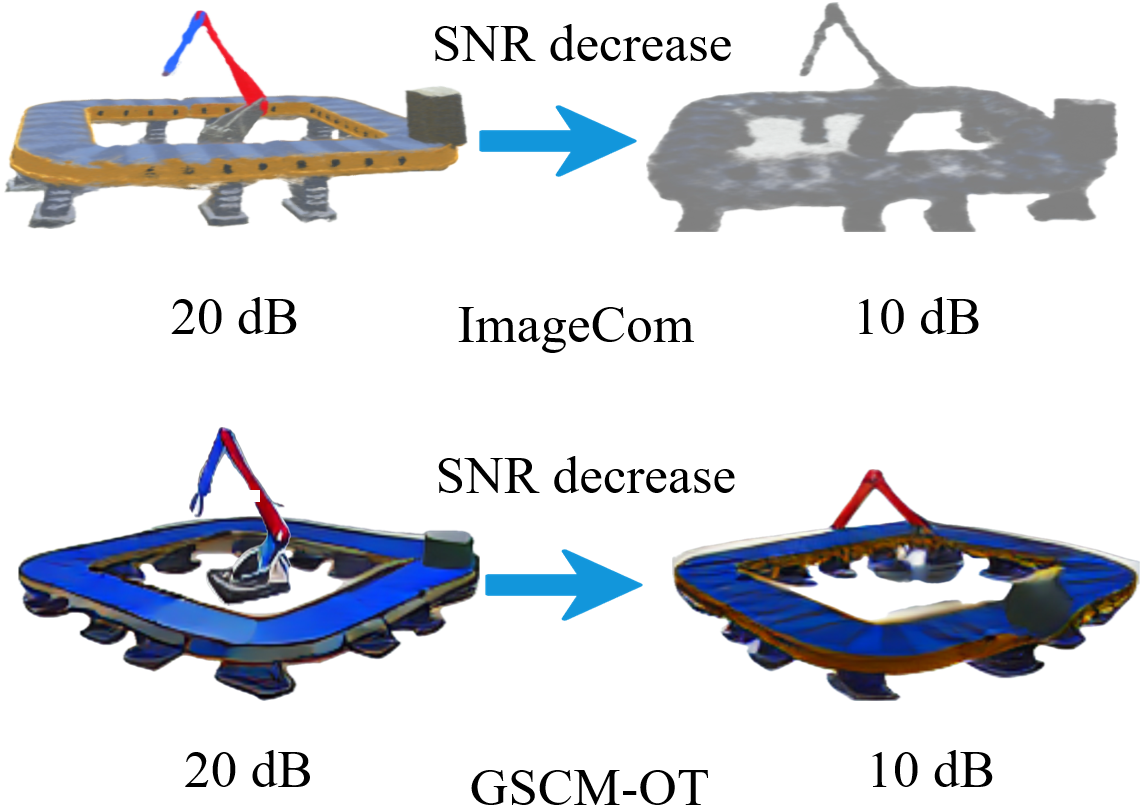} 
  \caption{Metaverse Construction with SNR Decrease} 
  \label{render_results} 
\end{figure}

For better evaluation, Fig. \ref{render_results} plots the results of metaverse construction under a Rayleigh channel. As the SNR decreases, both the proposed GSCM-OT and the ImageCom frameworks exhibit increased blurriness in the generated metaverse scenery. However, similar to the results discussed in Fig. \ref{exp2point}(c), we can observe that when the SNR is 20 dB, the robotic arm in the generated metaverse scene appears clearer, with the hook on the gripper distinctly visible. This is because rendering only the movable objects reduces the rendering space, allowing for a sharper depiction of the moving objects.
At 10 dB conditions, however, the metaverse scene generated by the ImageCom framework shows significant blurring, making it difficult to discern details. In contrast, while GSCM-OT also introduces some blurriness, the knowledge base allows for a relatively clearer background, including elements such as the conveyor belt. This demonstrates the enhanced transmission reliability that semantic communication provides in low SNR conditions.

\begin{figure}[t]
  \centering
  \includegraphics[width=0.42\textwidth]{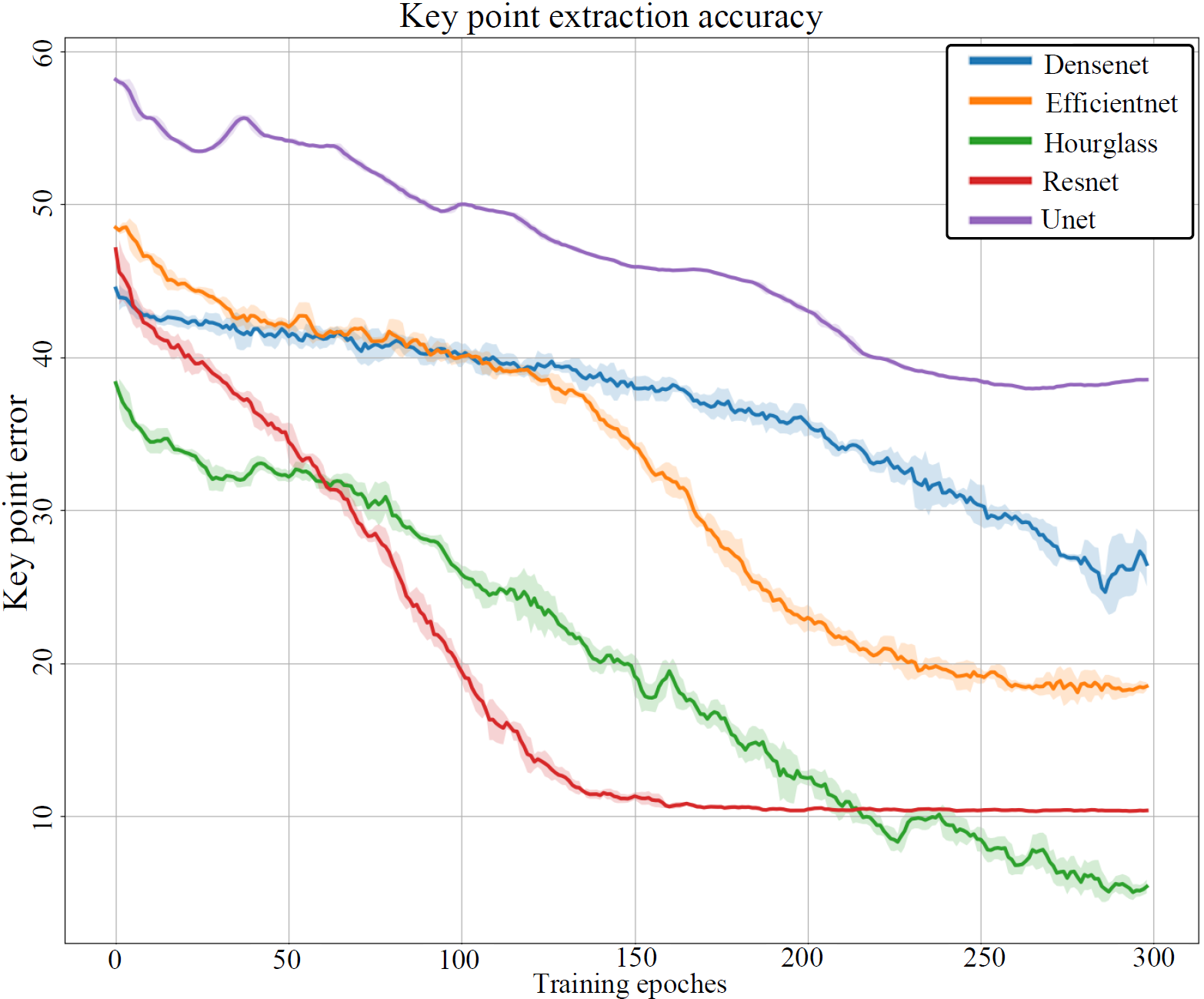} 
  \caption{Key Points Extraction Accuracy} 
  \label{KP_detection_Accruracy} 
\end{figure}

\subsection{Key Point Detection Accuracy}
Fig. \ref{KP_detection_Accruracy} plots the key points extraction by the HgNet semantic encoder, anchored to various backbone networks, including Hourglass, UNet, ResNet, and DenseNet, all evaluated over the same training epochs. Each neural network demonstrates a good ability to extract robotic arm and moving box key points from different images. The extraction accuracy performance with different backbone neural networks after 300 epochs is ranked as Hourglass $\textgreater$ PointConv $\textgreater$ RsNet $\textgreater$ EfficientNet.
The performance highlights the superior location prediction ability of the Hourglass-based semantic encoder in keypoint detection, achieving impressive accuracy with a margin of less than 3 pixels within the same training epoch's duration. Additionally, the results suggest that other backbone networks, such as ResNet, may perform more efficiently in feature extraction during the initial 100 epochs but ultimately exhibit lower precision in keypoint extraction. These networks may struggle with adequately extracting and learning from the structure of images unless the neural network architecture is deepened or made more complex, potentially impacting the accuracy of semantic information extraction. These findings underscore the significance of not only HgNet's key point extraction ability but also the careful selection of the backbone network when performing semantic information extraction on different data types.


\begin{figure}[t]
  \centering
  \includegraphics[width=0.42\textwidth]{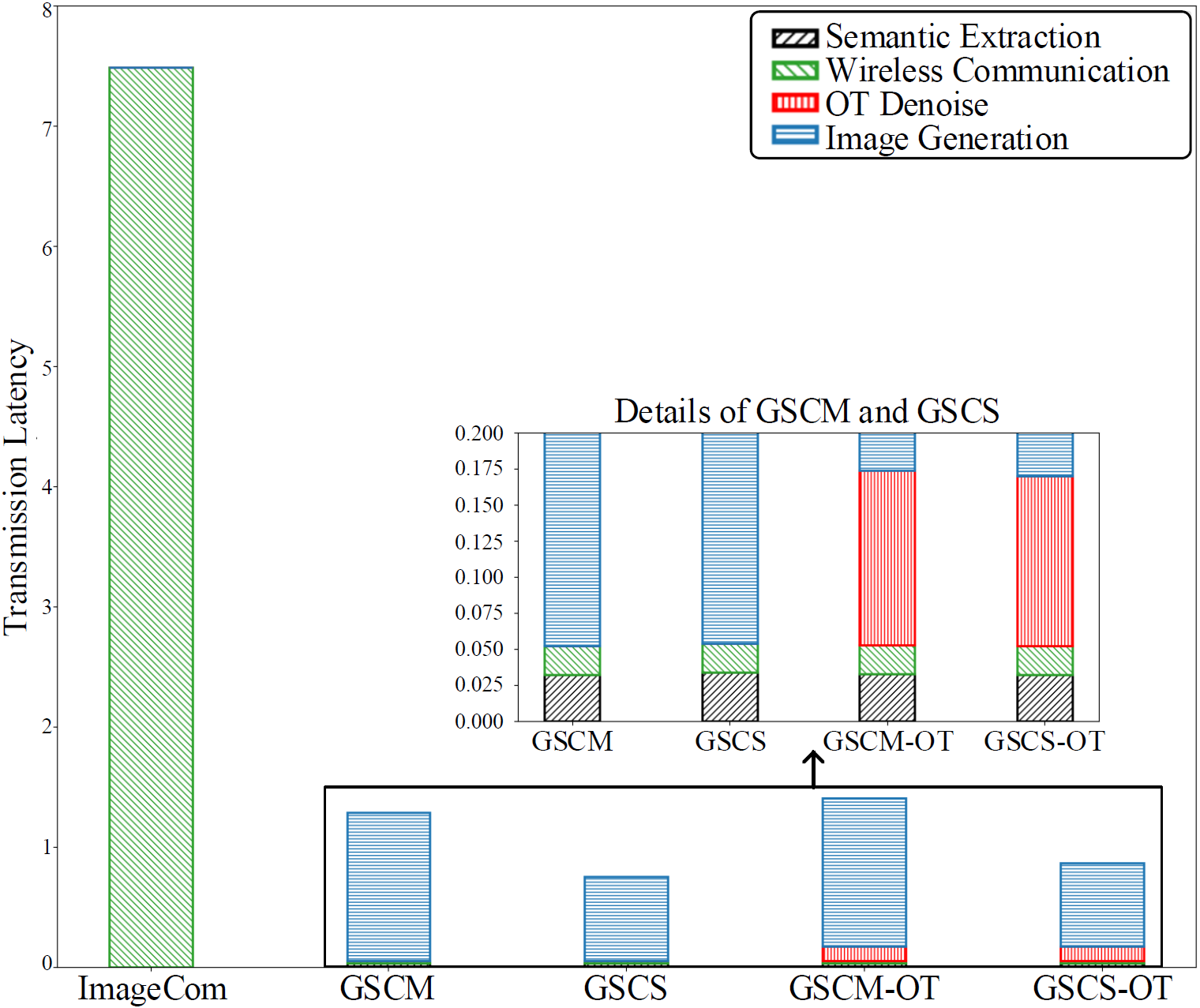} 
  \caption{Transmission Latency} 
  \label{time_latency} 
\end{figure}
\subsection{Transmission Latency}
Fig. \ref{time_latency} plots the transmission latency of all frameworks as defined in Eq. (19). Lower latency contributes to reduced metaverse construction time on the receiver side, with the frameworks ranked as follows: GSCS$\textless$GSCS-OT$\textless$GSCM$\textless$GSCM-OT$\textless$ImageCom. 
Since the NeRF construction latency appears in all the frameworks, the transmission delay here includes semantic extraction, wireless communication, OT denoising, and image generation.
Compared to the ImageCom framework, the proposed GSCM and GSCS frameworks significantly reduce transmission time due to the smaller amount of data transmitted. Although these frameworks introduce additional steps like semantic extraction, OT denoising, and image generation, these steps collectively take less than two seconds per frame, accounting for only a small fraction of the total transmission time.
Regarding image generation, which depends on receiver computing power and wireless bandwidth, the ImageCom framework requires receiving all images on the receiver side. In contrast, the GSCS and GSCM frameworks generate high-resolution images using a ControlNet-enabled Stable Diffusion algorithm, which only requires nine key points per image. With the aid of a powerful GPU, this approach drastically reduces the time required for receiving data on the receiver side.
As for OT-enabled denoising, each frame requires only 0.03 seconds for denoising, which constitutes a minimal part of the overall process compared to the performance gains from P2Point and KPE enhancements.
Specifically, the GSCM-OT framework achieves an 81.4\% reduction in transmission latency, while the GSCS-OT framework also achieves an 92.6\% reduction compared to the ImageCom framework. These improvements demonstrate the effectiveness of leveraging semantic information and optimized image generation algorithms to enhance the real-time performance of wireless metaverse applications, resulting in a smoother and more responsive user experience.

\section{Conclusion}
This paper proposed a goal-oriented semantic communication framework (GSC) to address the challenges of real-time communication and virtual world creation in the metaverse, particularly focusing on reducing latency and enhancing the accuracy of semantic information for virtual entities. By incorporating semantic information with the Neural Radiance Fields (NeRF) algorithm, the GSC framework selectively transmitted key semantic data, offering a more effective approach to information extraction than traditional communication frameworks, with fewer errors and reduced bandwidth requirements after wireless communication. Additionally, we implemented the Optimal Transport algorithm across varying wireless channel conditions within an end-to-end communication setup, distinguishing our approach and enhancing the general capabilities of semantic communication frameworks.
Our future work will optimize large-scale metaverse scenarios like universities and factories, and enhance metaverse construction by comparing the GSC framework under different wireless channels using machine learning for CSI feedback.

\vspace{-7pt}
\ifCLASSOPTIONcaptionsoff
  \newpage
\fi




\small
\bibliographystyle{IEEEtran}
\bibliography{IEEEabrv,Bibliography}

\vfill


\end{document}